\documentclass[11pt]{article}

\usepackage{fullpage}
\usepackage{epsfig}
\usepackage{amsfonts}
\usepackage{amssymb}
\usepackage{amstext}
\usepackage{amsmath}
\usepackage{xspace}
\usepackage{shadow,shadethm}
\usepackage{url}
\usepackage{times}

\makeatletter
\allowdisplaybreaks
\makeatother

\newtheorem{thm}{Theorem}
\newtheorem{lem}{Lemma}
\newtheorem{cl}{Claim}

\newcommand\ignore[1]{}

\newenvironment{pf}
{\noindent{\bf Proof:}}
{\hfill\rule{2mm}{2mm}\medskip}

\newenvironment{proofof}[1]
{\noindent{\bf Proof of {#1}:}}
{\hfill\rule{2mm}{2mm}\medskip}

\def\sse{\subseteq}
\def\opt{{\sf OPT}\xspace}
\def\alg{{\sf ALG}}
\def\is{{\mathbf{I}}}

\def\E{\mathbb{E}}

\def\lp{\ensuremath{{\sf LP}}\xspace}

\newcommand{\initOneLiners}{%
    \setlength{\itemsep}{0pt}
    \setlength{\parsep }{0pt}
    \setlength{\topsep }{0pt}
}
\newenvironment{OneLiners}[1][\ensuremath{\bullet}]
    {\begin{list} {#1} {\initOneLiners}} {\end{list}}

\usepackage{enumerate}

\newcounter{lpnumber} 
\newenvironment{linearprogram}[1]
{ \small \stepcounter{lpnumber}
  \begin{gather} {\textstyle #1} \tag{LP\arabic{lpnumber}} \\[-4ex] \notag
  \end{gather}
  \hspace{1.5cm} subject to \\[-3ex]
  \align }
{ \endalign }

\newcommand{\maximize}[1]{\text{maximize} \ #1}

\newcommand{\set}[1]{\left\{ #1 \right\}}
\newcommand{\sset}[2]{\left\{ #1 : #2 \right\}}
\newcommand{\abs}[1]{\left| #1 \right|}
\newcommand{\etal}{\textit{et~al.}}
\newcommand{\ceil}[1]{\left\lceil #1 \right\rceil}
\newcommand{\floor}[1]{\left\lfloor #1 \right\rfloor}
\newcommand{\assign}{\ensuremath{\leftarrow}}

\renewcommand{\deg}{\mathsf{deg}}

\newcommand{\pmax}{p_{\max}}

\newcommand{\hE}{\widehat{E}}

\renewcommand{\d}{\,\mathrm{d}}
\newcommand{\algo}{\mathsf{ALG}}

\title{When LP is the Cure for Your Matching Woes: \\ Improved Bounds for Stochastic Matchings}
\author{
Nikhil Bansal
\thanks{IBM T.J. Watson Research
Center, Yorktown Heights, NY 10598, USA.}
\hspace*{0.5in}
Anupam Gupta
\thanks{Computer Science Department, Carnegie Mellon
    University, Pittsburgh, PA 15213, USA.}
\hspace*{0.5in}
Jian Li
\thanks{
Computer Science Department. University of Maryland College Park, MD, USA.}
\hspace*{0.5in}
Juli\'{a}n Mestre
\thanks{
Max-Planck-Institut f\"{u}r
Informatik, Saarbr\"{u}cken, Germany.}
\\
Viswanath Nagarajan\footnotemark[1]
\hspace*{0.5in}
Atri Rudra
\thanks{
University at Buffalo, SUNY, Buffalo NY
14260, USA.
}
}

\date{}
\begin{document}
\maketitle

\begin{abstract}
  Consider a random graph model where each possible edge $e$ is
  present independently with some probability $p_e$. Given these
  probabilities, we want to build a large/heavy matching in the
  randomly generated graph.  However, the only way we can find out
  whether an edge is present or not is to query it, and if the edge is
  indeed present in the graph, we are forced to add it to our
  matching.  Further, each vertex $i$ is allowed to be queried at most
  $t_i$ times.  How should we \emph{adaptively} query the edges to
  maximize the expected weight of the matching? We consider several
  matching problems in this general framework (some of which arise in
  kidney exchanges and online dating, and others arise in modeling
  online advertisements); we give LP-rounding based constant-factor
  approximation algorithms for these problems. Our main results are
  the following:

\begin{OneLiners}
  \item We give a 4 approximation for weighted stochastic matching
  on general graphs, and a 3 approximation on bipartite
  graphs. This answers an open question from [Chen~\etal~ICALP 09].

  \item Combining our LP-rounding algorithm with the natural greedy
  algorithm, we give an improved 3.46 approximation for unweighted
  stochastic matching on general graphs.

  \item We introduce a generalization of the stochastic {\em online}
  matching problem \mbox{[Feldman~\etal~FOCS 09]} that also models
  preference-uncertainty and timeouts of buyers, and give a constant
  factor approximation algorithm.
\end{OneLiners}
\end{abstract}

\section{Introduction}

Motivated by applications in kidney exchanges and online dating, Chen~\etal~\cite{c-etal} proposed the following
stochastic matching problem: we want to find a maximum matching in a random graph $G$ on $n$ nodes, where each edge
$(i,j) \in [\binom{n}{2}]$ exists with probability $p_{ij}$, independently of the other edges. However, all we are
given are the probability values $\{p_{ij}\}$. To find out whether the random graph $G$ has the edge $(i,j)$ or not, we
have to try to add the edge $(i,j)$ to our current matching (assuming that $i$ and $j$ are both unmatched in our
current partial matching)---we call this ``probing'' edge $(i,j)$. As a result of the probe, we also find out if
$(i,j)$ exists or not---and if the edge $(i,j)$ indeed exists in the random graph $G$, it gets irrevocably added to
$M$. Such policies make sense, e.g., for dating agencies, where the only way to find out if two people are actually
compatible is to send them on a date; moreover, if they do turn out to be compatible, then it makes sense to match them
to each other. Finally, to model the fact that there might be a limit on the number of unsuccessful dates a person
might be willing to participate in, ``timeouts" on vertices are also provided. More precisely,  valid policies are
allowed, for each vertex $i$, to only probe at most $t_i$ edges incident to $i$. Similar considerations
arise in kidney exchanges, details of which appear in~\cite{c-etal}.

Chen~\etal\ asked the question: how can we devise probing policies to
maximize the expected cardinality (or weight) of the matching?
They showed that the greedy algorithm that probes edges in decreasing
order of $p_{ij}$ (as long as their endpoints had not timed out) was a
$4$-approximation to the cardinality version of the stochastic matching
problem. This greedy algorithm (and other simple greedy schemes) can be
seen to be arbitrarily bad in the presence of weights, and they left
open the question of obtaining good algorithms to \emph{maximize the
  expected weight} of the matching produced. In addition to being a
natural generalization, weights can be used as a proxy for revenue
generated in matchmaking services. (The unweighted case
can be thought of as maximizing the social welfare.) In this paper, we
resolve the main open question from Chen~\etal~\cite{c-etal}:
\begin{thm}
  \label{th:main1}
  There is a $4$-approximation algorithm for the weighted  stochastic matching problem.
  For bipartite graphs, there is a $3$-approximation algorithm.
\end{thm}

Our main idea is to use the knowledge of edge probabilities to solve a linear program where each edge $e$ has a
variable $0\le y_e\le 1$ corresponding to the probability that a strategy probes $e$ (over all possible realizations of
the graph). This is similar to the approach for stochastic packing problems considered by Dean et
al.~\cite{dgv04,dgv05}. We then give two different rounding procedures to attain the bounds claimed above.
\begin{OneLiners}
 \item The first algorithm (\S\ref{subsec:wt-match}) is very simple: it considers edges in a uniformly random order
and probes each edge $e$ with probability proportional to $y_e$; the analysis uses Markov's inequality and a
Chernoff-type bound (Lemma~\ref{lem:high-TO-prob}).
 \item The second algorithm (\S\ref{subsec:wt-bip}) is more nuanced and
achieves a better bound: we use {\em dependent rounding}~\cite{gandhi2006dependent} on the $y$-values to obtain a set
$\hat{E}$ of edges to be probed, and then probe edges of $\hat{E}$ in a uniformly random order.
\end{OneLiners}
Though the first algorithm has a weaker approximation ratio, we still present it since it is useful in the online
stochastic matching problem (Section~\ref{sec:online}).


The second rounding algorithm has an additional advantage: The probing strategy returned by the algorithm can be made
{\em matching-probing}~\cite{c-etal}. In this alternative (more restrictive) probing model we are given an additional
parameter $k$ and edges need to be probed in $k$ rounds, each round being a matching. It is clear that this
matching-probing model is more restrictive than the usual {\em
  edge-probing} model (with timeouts $\min\{t_i,k\}$) where one edge
is probed at a time. Our algorithm obtains a matching-probing strategy
that is only a small constant factor worse than the optimal
edge-probing strategy; hence, we also obtain the same constant
approximation guarantee for weighted stochastic matching in the
matching-probing model. It is worth noting that previously only a
logarithmic approximation in the unweighted case was
known~\cite{c-etal}.

\begin{thm}
  \label{th:main1'}
  There is a $4$-approximation algorithm for the weighted stochastic
  matching problem in the {\bf matching-probing} model.  For bipartite
  graphs, there is a $3$-approximation algorithm.
\end{thm}

Notice that for general graphs our algorithm matches the performance
of the greedy algorithm shown by Chen~\etal~\cite{c-etal} for the
unweighted case. Interestingly, even though their individual analyses
show that they are 4-approximations, they can be combined to obtain
better approximations.

\begin{thm}
  \label{th:main1.5}
  There is a $3.46$-approximation algorithm for the unweighted
  stochastic matching problem in general graphs.
\end{thm}

Apart from solving these open problems and yielding improved
approximations, our LP-based analysis turns out to be applicable in a
wider context.

\paragraph{{\em Online} Stochastic Matching Revisited.} In a bipartite
graph $(A,B;E)$ of items $i \in A$ and potential buyer types $j \in
B$, $p_{ij}$ denotes the probability that a buyer of type $j$ will buy
item $i$.  A sequence of $n$ buyers are to arrive online, where the
type of each buyer is an i.i.d.\ sample from $B$ according to some
pre-specified distribution---when a buyer of type $j$ appears, he can
be shown a list $L$ of up to $t_j$ as-yet-unsold items, and the buyer
buys the \emph{first} item on the list according to the given
probabilities $p_{\cdot, j}$.  (Note that with probability $\prod_{i
  \in L} (1 - p_{ij})$, the buyer leaves without buying anything.)
What items should we show buyers when they arrive online, and in which order, to maximize the expected weight of the
matching? Building on the algorithm for stochastic matching in~\S\ref{subsec:wt-gen}, we prove the following in
Section~\ref{sec:online}.

\begin{thm}
  \label{th:main3}
  There is a $7.92$-approximation algorithm for the above online stochastic
  matching problem.
\end{thm}

This question is an extension of similar online stochastic matching
questions considered earlier in~\cite{FMMM}---in that paper, $w_{ij},
p_{ij} \in \{0,1\}$ and $t_j = 1$. Our model tries to capture the
facts that buyers may have a limited attention span (using the
timeouts), they might have uncertainties in their preferences (using
edge probabilities), and that they might buy the first item they like
rather than scanning the entire list.

\paragraph{A New Proof for Greedy.} The proof in~\cite{c-etal} that
the greedy algorithm for stochastic matching was a $4$-approximation in the unweighted case was based on a somewhat
delicate charging scheme involving the decision trees of the algorithm and the optimal solution.  We show
(Appendix~\ref{sec:unwt-greedy}) that the greedy algorithm, which was defined without reference to any LPs, admits a
simple LP-based analysis.
\begin{thm}
  \label{th:main2}
  The greedy algorithm is a $5$-approximation for the unweighted stochastic matching problem.
\end{thm}

\paragraph{Cardinality Constrained Matching in Rounds.} We also consider the model from~\cite{c-etal} where one
can probe as many as $C$ edges in parallel,  as long as these $C$ edges form a matching; the goal is to maximize
  the expected weight of the matched edges after $k$ rounds of such
  probes. We improve on the $\min\{k,C\}$-approximation offered
  in~\cite{c-etal} (which only works for the unweighted version), and show in Appendix~\ref{sec:mult-match}:
\begin{thm}
  \label{th:main4}
  There is a constant-factor approximation algorithm for
  weighted cardinality constrained multiple-round stochastic matching.
\end{thm}

\paragraph{Extension to Hypergraphs.} We extend our analysis to a much more general situation where we
  try to pack $k$-hyperedges with random sizes into a $d$-dimensional
  knapsack of a given size; this is just the stochastic knapsack problem
  of~\cite{dgv05}, but where we consider the situation where $k \ll d$.
  For this setting of parameters, we improve on the
  $\sqrt{d}$-approximation of~\cite{dgv05} to prove the following (Section~\ref{sec:set-packing}).
\begin{thm}
  \label{th:main5}
  There is a $2k$-approximation algorithm for the weighted stochastic
  $k$-set-packing problem.
\end{thm}

We note that the stochastic $k$-set-packing problem is a direct generalization of the stochastic matching problem; so
an 8-approximation for stochastic matching follows from Theorem~\ref{th:main5}. However, using more structure in the
matching problem, we could obtain the better approximation ratios in Theorems~\ref{th:main1} and~\ref{th:main1'}.


\subsection{Related Work.}

As mentioned above, perhaps the work most directly related to this
work is that on stochastic knapsack problems (Dean et
al.~\cite{dgv04,dgv05}) and multi-armed bandits (see~\cite{gm07,gm09}
and references therein).  Also related is some recent
work~\cite{bhattacharya-stoc10} on budget constrained auctions, which
uses similar LP rounding ideas.

In recent years stochastic optimization problems
have drawn much attention from the theoretical computer science community where
stochastic versions of several classical combinatorial optimization problems
have been studied.
Some general techniques have also been developed \cite{gupta2004boosted,shmoys2006approximation}.
See \cite{swamy2006approximation} for a survey.

The online bipartite matching problem was first studied in the seminal
paper by Karp \etal~\cite{KarpVV90} and an optimal $1-1/e$ competitive
online algorithm was obtained.  Katriel
\etal~\cite{katriel2008commitment} considered the two-stage stochastic
min-cost matching problem.  In their model, we are given in a first
stage probabilistic information about the graph and the cost of the
edges is low; in a second stage, the actual graph is revealed but the
costs are higher.
The original online stochastic matching problem was studied recently by
Feldman~\etal~\cite{FMMM}.  They gave a $0.67$-competitive algorithm,
beating the optimal $1-1/e$-competitiveness known for worst-case
models~\cite{KarpVV90,KalyanasundaramP93,MSVV05,BirnbaumM08,GoelM08}.
Our model differs from that in having a bound on the number of items
each incoming buyer sees, that each edge is only present with some
probability, and that the buyer scans the list linearly (until she times
out) and buys the first item she likes.
Recently, some improved bounds on this model were obtained \cite{bahmani10improved,manshadi10online}.

Our problem is also related to the Adwords problem \cite{MSVV05},
which has applications to sponsored search auctions.  The problem can be
modeled as a bipartite matching problem as follows. We want to
assign every vertex (a query word) on one side to a vertex (a bidder) on the
other side. Each edge has a weight, and there is a budget on each bidder
representing the upper bound on the total weight of edges that may be assigned
to it. The objective is to maximize the total revenue.
The stochastic version in which query words arrive according to
some known probability distribution has also been studied
\cite{mahdian2007allocating}.

\subsection{Preliminaries.}
\label{subsec:prelim} For any integer $m\ge 1$, define $[m]$ to be the set $\{1,\dots,m\}$. For a maximization problem,
an $\alpha$-approximation algorithm is one that computes a solution with expected objective value at least $1/\alpha$
times the expected value of the optimal solution.

We must clarify here the notion of an optimal solution.  In standard
worst case analysis we would compare our solution against the optimal
\textit{offline} solution, e.g. the value of the maximum matching,
where the offline knows all the edge instantiations in advance
(i.e. which edge will appear when probed, and which will not).
However, it can be easily verified that due to the presence of
timeouts, this adversary is too strong~\cite{c-etal}.  Consider the
following example.  Suppose we have a star where each vertex has
timeout 1, and each edge has $p_{ij}=1/n$. The offline optimum can
match an edge whenever the star has an edge i.e. with probability
about $1-1/e$, while our algorithm can only get expected $1/n$ profit,
as it can only probe a single edge.  Hence, for all problems in this
paper we consider the setting where even the optimum does not know the
exact instantiation of an edge until it is probed.  This gives our
algorithms a level playing field. The optimum thus corresponds to a
``strategy" of probing the edges, which can be chosen from an
exponentially large space of potentially adaptive strategies.

We note that our algorithms in fact yield {\em non-adaptive}
strategies for the corresponding problems, that are only constant
factor worse than the adaptive optimum. This is similar to previous
results on stochastic packing problems: knapsack
(Dean~\etal~\cite{dgv04,dgv05}) and multi-armed bandits
(Guha-Munagala~\cite{gm07,gm09} and references therein).

\section{Stochastic Matching}
\label{sec:stoc-match} We consider the following stochastic matching
problem. The input is an undirected graph $G=(V,E)$ with a weight
$w_e$ and a probability value $p_e$ on each edge $e\in E$. In
addition, there is an integer value $t_v$ for each vertex $v\in V$
(called \textit{patience parameter}). Initially, each vertex $v\in V$
has patience $t_v$. At each step in the algorithm, any edge $e(u,v)$
such that $u$ and $v$ have positive remaining patience can be
probed. Upon probing edge $e$, one of the following happens: (1) with
probability $p_{e}$, vertices $u$ and $v$ get {\em matched} and are
removed from the graph (along with all adjacent edges), or (2) with
probability $1-p_{e}$, the edge $e$ is removed and the remaining
patience numbers of $u$ and $v$ get reduced by $1$. An algorithm is an
adaptive strategy for probing edges: its performance is measured by
the expected weight of matched edges. The {\em unweighted} stochastic
matching problem is the special case when all edge-weights are
uniform.

Consider the following linear program: as usual, for any vertex $v\in
V$, $\partial(v)$ denotes the edges incident to $v$. Variable $y_{e}$
denotes the probability that edge $e=(u,v)$ gets probed in the adaptive
strategy, and $x_{e}=p_{e}\cdot y_{e}$ denotes the probability that
$u$ and $v$ get matched in the strategy. (This LP is similar to the LP
used for general stochastic packing problems by Dean, Goemans and
Vondr{\'a}k~\cite{dgv05}.)
\begin{linearprogram}
  {
    \label{LP:stoc-matching}
    \maximize{\displaystyle \sum_{e\in E} w_{e}\cdot x_{e}}
  }
  \sum_{e\in \partial(v)} x_{e} &\le 1&& \forall v\in V  \label{eq:lp2}\\
  \sum_{e\in \partial(v)} y_{e} & \le t_i&& \forall v\in V \label{eq:lp3}\\
  x_{e} &=p_{e}\cdot y_{e} && \forall e\in E \label{eq:lp4}\\
  0\le y_{e} &\le 1 &&\forall e\in E \label{eq:lp5}
\end{linearprogram}

The following claim shows that the LP above is a valid relaxation for the
stochastic matching problem.

\begin{cl}
  \label{clm:adaptive}
  The optimal value for  LP~\eqref{LP:stoc-matching} is an upper bound on any (adaptive)
  algorithm for stochastic matching.
\end{cl}

\begin{pf}
  To show this, it suffices to show that any adaptive strategy
  satisfies the constraints of the LP.  Conditioned on any instantiation of all
  edges in $E$ (i.e. each edge $e\in E$ is present with probability
  $p_e$), the expected number of probes involving any vertex $v\in V$
  is at most $t_v$ (the patience parameter).  Similarly conditioning on
  edges $E$, the expected number of matched edges involving $v\in V$ is
  at most $1$. Hence these constraints hold unconditionally as well, which implies
that any valid strategy satisfies (\ref{eq:lp2}) and (\ref{eq:lp3}).
\end{pf}


\subsection{Weighted Stochastic Matching: General
  Graphs}
\label{subsec:wt-match}
Our algorithm first
solves \eqref{LP:stoc-matching} to optimality and uses the optimal
solution $(x,y)$ to obtain a non-adaptive strategy achieving expected
value $\Omega(1)\cdot (w\cdot x)$. Next, we present the algorithm. Let
$(x,y)$ denote an optimal solution to the above LP, which by
Claim~\ref{clm:adaptive} gives an upper-bound on any adaptive
strategy.  Let $\alpha\ge 1$ be a constant to be set later.  The
algorithm first fixes a uniformly random permutation $\pi$ on edges
$E$. It then inspects edges in the order of $\pi$, and {\em probes}
only a subset of the edges. A vertex $v\in V$ is said to have {\em
  timed out} if $t_v$ edges incident to $v$ have already been probed
(i.e. its remaining patience reduces to 0); and vertex $v$ is said to
be {\em matched} if it has already been matched to another vertex. An
edge $(u,v)$ is called {\em safe} at the time it is considered if (A)
neither $u$ nor $v$ is matched, and (B) neither $u$ nor $v$ has timed
out. The algorithm is the following:

\begin{shadebox}
  \begin{OneLiners}
  \item[1.] Pick a permutation $\pi$ on edges $E$ uniformly at random
  \item[2.] For each edge $e$ in the ordering $\pi$, do:
    \begin{OneLiners}
    \item[a.] If $e$ is safe then probe it with probability
      $y_{e}/\alpha$, else do not probe it.
    \end{OneLiners}
  \end{OneLiners}
\end{shadebox}

In the rest of this section, we prove that this algorithm achieves a
$5.75$-approximation for the weighted stochastic matching
problem. Even though this is slightly worse that the approximation
factors claimed in Theorem~\ref{th:main1}, this first algorithm is
significantly simpler, it readily illustrates the power of the LP
approach, and, as we shall see in \S~\ref{sec:set-packing}, it can
handle a much more general version of the basic problem.

We begin with the following property:

\begin{lem}\label{lem:match-prob}
  For any edge $(u,v)\in E$, at the point when $(u,v)$ is considered
  under $\pi$,
  \begin{OneLiners}
  \item[(a)] the probability that vertex $u$ has timed out is at most
    $\frac{1}{2\alpha}$, and
  \item[(b)] the probability that vertex $u$ is matched is at most
    $\frac{1}{2\alpha}$.
  \end{OneLiners}
\end{lem}

\begin{pf}
  We begin with the proof of part (a). Let
  random variable $U$ denote the number of probes incident to vertex $u$
  by the time edge $(u,v)$ is considered in $\pi$.
  \begin{align*}
    \E[U]&= \sum_{e\in \partial(u)} \Pr [\, \mbox{edge $e$ appears before
      $(u,v)$ in $\pi$ AND $e$ is probed}\, ], \\
    &\le \sum_{e\in \partial(u)} \Pr [ \, \mbox{edge $e$ appears before $(u,v)$
      in $\pi$}\, ]\cdot \frac{y_e}\alpha, \\
    & = \sum_{e\in \partial(u)} \frac{y_e}{2\alpha}, \\
    & \le \frac{t_u}{2\alpha}.
  \end{align*}
  The first inequality above follows from the fact that the
  probability that edge $e$ is probed (conditioned on $\pi$) is at
  most $y_e/\alpha$. The second equality follows since $\pi$ is a
  u.a.r.  permutation on $E$. The last inequality is by the LP
  constraint~\eqref{eq:lp3}. The probability that vertex $u$ has timed
  out when $(u,v)$ is considered equals $\Pr[\, U\ge t_u \, ] \le
  \frac{\E[U]}{t_u}\le \frac1{2\alpha}$, by the Markov
  inequality. This proves part (a). The proof of part (b) is identical
  (where we consider the event that an edge is matched instead of
  being probed and replace $y_e$ and $t_u$ by $x_e$ and $1$
  respectively and use the LP constraint (\ref{eq:lp2})) and is
  omitted.
\end{pf}

Now, a vertex $u\in V$ is called {\em low-timeout} if $t_u=1$, else
$u$ is called a {\em high-timeout} vertex if $t_u\ge 2$. We next prove
the following bound for high-timeout vertices that is stronger than
the one from Lemma~\ref{lem:match-prob}(a).
\begin{lem}\label{lem:high-TO-prob}
  Suppose $\alpha\ge \mathrm{e}$. For a high-timeout vertex $u\in V$, and any
  edge $f$ incident to $u$, the probability that $u$ has timed out when
  $f$ is considered in $\pi$ is at most $\frac2{3\alpha^2}$.
\end{lem}
\begin{pf}
  Let $t=t_u\ge 2$ denote the patience parameter for vertex $u$, and
  $F=\partial(u)\setminus \{f\}$ the set of edges incident to $u$
  excluding $f$. Then the probability that $u$ has timed out when $f$
  is considered under $\pi$ is upper bounded by:
  \begin{align}
    \sum_{\{p_1,\cdots,p_t\}\sse F} & \Pr[ \, \mbox{edges $p_1,\cdots,p_t$ appear before $f$ in $\pi$ AND are all probed}\,], \label{eq:TO1}\\
&\le \frac1{t!} \cdot \sum_{p_1,\cdots,p_t\in F} \Pr[ \, \mbox{edges
  $p_1,\cdots,p_t$ appear before $f$ in $\pi$ AND are all probed}\, ], \label{eq:TO2}\\
&\le \frac1{t!} \cdot \sum_{p_1,\cdots,p_t\in F} \Pr [ \, \mbox{edges
  $p_1,\cdots,p_t$ appear before $f$ in $\pi$}\, ] \cdot
\prod_{\ell=1}^t \frac{y_{p_{\ell}}}{\alpha}, \label{eq:TO3}\\
&= \frac1{(t+1)!} \cdot \sum_{p_1,\cdots,p_t\in F} \prod_{\ell=1}^t \frac{y_{p_{\ell}}}{\alpha}, \label{eq:TO4}\\
&=  \frac1{(t+1)!} \cdot \left(\,\sum_{p\in F} \frac{y_p}{\alpha}\,\right)^t\\
&\le  \frac1{(t+1)!} \cdot\left(\frac{t}{\alpha}\right)^t. \label{eq:TO6}
\end{align}
In the above, the summation in~\eqref{eq:TO1} is over unordered $t$-tuples
whereas the subsequent ones~\eqref{eq:TO2}-\eqref{eq:TO4} are over
ordered tuples (with repetition). Inequality~\eqref{eq:TO3} uses the
fact that for any edge $g$, the probability of probing $g$ {\em
  conditioned} on $\pi$ and the outcomes until $g$ is considered, is at
most $y_g/\alpha$ (and the fact that the probability of probing an edge is independent
of the probability of probing other edges). Equation~\eqref{eq:TO4} follows from the fact that
probability that $f$ is the last to appear among $\{p_1,\cdots,p_t,f\}$
in a random permutation $\pi$ is $\frac1{t+1}$. Finally,~\eqref{eq:TO6}
follows from the LP constraint~\eqref{eq:lp3} at $u$.

Let $f(t):=\frac1{(t+1)!} \cdot\left(\frac{t}{\alpha}\right)^t$. We
claim that $f(t)\le \frac2{3\alpha^2}$ when $\alpha\ge \mathrm{e}$ and $t\ge 2$,
which would prove the claim. Note that this is indeed true for $t=2$ (in
fact with equality). Also $f(t+1)\le f(t)$ for all $t\ge 2$ due to:
$$\frac{f(t+1)}{f(t)} = \left(\frac{t+1}t\right)^t\cdot \frac{t+1}{t+2}\cdot \frac{1}{\alpha} \le \frac{\mathrm{e}}{\alpha}\le 1.$$
Thus we obtain the desired upper bound.
\end{pf}

Using this, we can analyze the probability that an edge is safe.
\begin{lem}
  \label{lem:safe-prob}
  For $\alpha \geq \mathrm{e}$, an edge $f=(u,v)$ is safe with probability
  at least $(1-\frac1\alpha -\frac4{3\alpha^2})$ when $f$ is considered
  under a random permutation $\pi$.
\end{lem}
\begin{pf}
  The analysis proceeds by considering the following cases.
  \begin{enumerate}
  \item \label{case:date1} {\em Both $u$ and $v$ are low-timeout.} Since
    $t_u=t_v=1$, the event that $u$ (resp. $v$) is matched at any point
    is a subset of the event that $u$ (resp. $v$) has timed out.  Thus
    by Lemma~\ref{lem:match-prob}, the probability that edge $f$
    is not safe (when it is considered) is  $\leq\frac{2}{2\alpha}$.
  \item \label{case:date2} {\em Both $u$ and $v$ are high-timeout.}
    Lemma~\ref{lem:high-TO-prob} implies that the probability that
    $u$ (resp. $v$) has timed out is at most $\frac2{3\alpha^2}$. Again
    by Lemma~\ref{lem:match-prob}, the probability that $u$ (resp.
    $v$) is matched is at most $\frac1{2\alpha}$.  Thus the probability
    that $f$ is not safe is at most $\frac1\alpha
    +\frac4{3\alpha^2}$.
  \item {\em $u$ is low-timeout and $v$ is high-timeout.} Using the
    argument in Step~\eqref{case:date1} for vertex $u$, the probability
    that vertex $u$ has timed out or matched is at most
    $\frac1{2\alpha}$. And using Step~\eqref{case:date2} for vertex $v$,
    the probability that vertex $v$ has timed out or matched is at most
    $\frac1{2\alpha}+\frac2{3\alpha^2}$. So the probability that edge
    $(u,v)$ is not safe is at most $\frac1\alpha +\frac2{3\alpha^2}$.
  \end{enumerate}
  Hence every edge is safe (when considered in $\pi$) with
  probability $\ge (1-\frac1\alpha -\frac4{3\alpha^2})$.
\end{pf}

\begin{thm}
  \label{thm:wtd}
  Setting $\alpha = 1+\sqrt{5}$ in the above algorithm gives an
  $5.75$-approximation for the weighted stochastic matching problem.
\end{thm}

\begin{pf}{Theorem~\ref{th:main1}}
  Given that an edge $e\in E$ is safe when considered, the expected
  profit for the algorithm is $w_e\cdot p_e\,\frac{y_e}\alpha=w_e\cdot
  x_e/\alpha$. Now using Lemma~\ref{lem:safe-prob}, the algorithm gets
  expected profit at least $(\frac1\alpha
  -\frac1{\alpha^2}-\frac4{3\alpha^3})$ times the optimal LP value.
  Plugging in $\alpha = 1+\sqrt{5} $ gives an approximation ratio of
$\frac{3(16+8\sqrt{5})}{11+3\sqrt{5}}<5.75$, as desired.
\end{pf}


\subsection{Weighted Stochastic Matching: Bipartite Graphs}
\label{subsec:wt-bip} In this section, we obtain an improved bound for
stochastic matching on bipartite graphs via a different rounding
procedure. In fact, the algorithm produces a {\em matching-probing
  strategy} whose expected value is a constant fraction of the optimal
value of \eqref{LP:stoc-matching} (which was for edge-probing).

\paragraph{Algorithm.} First, we find an optimal fractional solution
$(x,y)$ to \eqref{LP:stoc-matching} and round $y$ to identify a set of
interesting edges $\widehat{E}$. Then we use K\"onig's Theorem
\cite[Ch. 20]{book/combopt/Schrijver} to partition $\widehat{E}$ into
a small collection of matchings $M_1, \dots, M_h$. Finally, these
matchings are then probed in random order. If we are only interested
in edge-probing strategies, probing the edges in $\widehat{E}$ in
random order would suffice. We will refer to this algorithm as {\sc
  round-color-probe}:

\begin{shadebox}
  \begin{OneLiners}
    \item[1.] $(x,y)$ \assign\ optimal solution to \eqref{LP:stoc-matching}
    \item[2.] $\widehat{y}$ \assign\ round $y$ to an integral
    solution \label{line:rounding} using GKSP
    \item[3.] $\hE \assign \sset{ e \in E }{ \widehat{y}_e =1 }$
    \item [4.] $M_1, \ldots, M_h$ \assign\ optimal edge coloring of
    $\hE$
    \item[5.] For each $M$ in $\set{M_1, \ldots, M_h}$ in
    random order, do:
    \begin{OneLiners}
      \item[a.] probe $\sset{ (u,v) \in M}{ \text{$u$ and $v$
          are unmatched} }$
    \end{OneLiners}
  \end{OneLiners}
\end{shadebox}

Besides the edge coloring step, the key difference from the algorithm
of the previous subsection is in the choice of $\widehat{E}$. For this
we use the GKSP procedure of Gandhi~\etal~\cite{gandhi2006dependent},
which we describe next.

\paragraph {The GKSP algorithm.} We state some properties
of the dependent rounding framework of
Gandhi~\etal~\cite{gandhi2006dependent} that are relevant in our
context.

\begin{thm}[\cite{gandhi2006dependent}]
  \label{thm:dep-rounding}
  Let $(A, B; E)$ be a bipartite graph and $z_{e} \in [0,1]$ be
  fractional values for each edge $e \in E$. The GKSP algorithm is a
   polynomial-time randomized procedure that outputs values $Z_e \in
  \{0,1\}$ for each $e \in E$ such that the following properties hold:

\begin{enumerate}
  \item[P1.] {\it Marginal distribution.} For every edge $e$, $\Pr[Z_e = 1] = z_e$.
  \item[P2.] {\it Degree preservation.}  For every vertex $u \in A
  \cup B$,  $\sum_{e\in \partial(u)} Z_{e} \leq  \ceil{\sum_{e\in \partial u} z_e}$.
  \item[P3.] {\it Negative correlation.}  For any vertex $u$ and any set of
  edges $S \subseteq{\partial(u)}$:
  \begin{equation*}
    \Pr[\bigwedge_{e \in S} (Z_e = 1)] \leq \prod_{e \in S} \Pr[Z_e = 1].
  \end{equation*}
\end{enumerate}
\end{thm}

We note that the GKSP algorithm in fact guarantees stronger properties
than the ones stated above. For the purpose of analyzing {\sc
  round-color-probe}, however, the properties stated above will
suffice.

\paragraph{Feasibility.} Let us first argue that our algorithm outputs
a feasible strategy. If we care about feasibility in the
edge-probing model, we only need to show that each vertex $u$ is not
probed more than $t_u$ times. The following lemma shows that:

\begin{lem}
  \label{lem:rcp-feasible}
  For every vertex $u$, {\sc round-color-probe} probes at most $t_u$
  edges incident on $u$.
\end{lem}

\begin{pf}
  Vertex $u$ is matched in $\abs{ \set{e
      \in \partial_{\widehat{E}}(u)}}$ matchings. This is an upper
  bound on the number of times edges incident on $u$ probed. Hence we
  just need to show that this quantity is at most $t_u$. Indeed,
  \[\abs{\set{e \in \partial_{\widehat{E}}(u)}} = \sum_{e
    \in \partial(u)} \hat{y}_e \leq \Big\lceil \! \sum_{e
      \in \partial(u)} y_e \Big\rceil \leq t_u, \]
  where the first inequality follows from the degree preservation
  property of Theorem~\ref{thm:dep-rounding} and the second inequality
  from the fact that $y$ is a feasible solution to
  \eqref{LP:stoc-matching}.
\end{pf}

Let us argue that the strategy is also feasible under the
matching-probing model. Recall that in the latter model we are given
an additional parameter $k$ (which without loss of generality we can
assume to be at most $\max_{v \in V}t_u$) and we can probe edges in
$k$ round, with each round forming a matching. Let $\hE$ be the set of
edges in the support of $\widehat{y}$, i.e., $\hE=\{e \in E \mid
\widehat{y}_e=1\}$. Let $h = \max_{v \in V} \deg_{\hE}(v) \leq \max_{v
  \in V} t_v$. K\"{o}nig's Theorem allows us to decomposed $\hE$ into
$h$ matchings. Therefore, the probing strategy devised by the
algorithm is also feasible in the matching-probing model.

\paragraph{Performance guarantee.} Let us focus our attention on some
edge $e=(u,v) \in E$. Our goal is to show that there is good chance
that the algorithm will indeed probe $e$. We first analyze the
probability of $e$ being probed conditional on $\hE$. Notice that the
algorithm will probe $e$ if and only if all previous probes incident
on $u$ and $v$ were unsuccessful; otherwise, if there was a successful
probe incident on $u$ or $v$, we say that $e$ was \emph{blocked}.

Let $\pi$ be a permutation of the matchings $M_1, \ldots, M_h$. We
extend this ordering to the set $\hE$ by listing the edges within a
matching in some arbitrary but fixed order. Let us denote by $B(e,
\pi) \subseteq \widehat{E}$ the set of edges incident on $u$ or $v$
that appear before $e$ in $\pi$. It is not hard to see that
\begin{align}
  \label{eq:prob-probed}
  \Pr\, [\, e \text{ was not blocked}\,\mid\, \widehat{E} \,]
     & \geq \E_\pi\Bigr[\, \prod_{f \in B(e, \pi)}
       (1-p_f)\, \mid \, \widehat{E} \, \Bigr];
\end{align}
here we assume that $\prod_{f \in B(e, \pi)} (1-p_f)=1$ when $B(e,
\pi)=\emptyset$.

Notice that in~\eqref{eq:prob-probed} we only care about the order of
edges incident on $u$ and $v$. Furthermore, the expectation does not
range over all possible orderings of these edges, but only those that
are consistent with some matching permutation. We call this type of
restricted ordering \emph{random matching ordering} and we denote it
by $\pi$; similarly, we call an unrestricted ordering \emph{random
  edge ordering} and we denote it by $\sigma$. Our plan is to study
first the expectation in \eqref{eq:prob-probed} over random edge
orderings and then to show that the expectation can only increase when
restricted to range over random matching orderings.

The following simple lemma is useful in several places.

\begin{lem}
  \label{lem:eta} Let $r$ and $p_{max}$ be positive real
  values. Consider the problem of minimizing $\prod_{i=1}^t (1-p_i)$
  subject to the constraints $\sum_{i=1}^t p_i\leq r$ and $0\leq
  p_i\leq \pmax$ for $i=1,\ldots, t$. Denote the minimum value by
  $\eta(r,\pmax)$.  Then,
  \[ \eta(r,\pmax)= (1-\pmax)^{\floor{\frac{r}{\pmax}}} \left(1-(r-{\textstyle
    \floor{\frac{r}{\pmax}}} \pmax)\right) \geq (1-p_{\max})^{r/p_{\max}}.\]
\end{lem}

\begin{pf}
  Suppose the contrary that the quantity is minimized but there are two $p_i$s
  that are strictly between $0$ and $\pmax$. W.l.o.g, they are $p_1,p_2$ and
  $p_1>p_2$ Let $\epsilon=\min(\pmax-p_1, p_2)$.  It is easy to see that
  \begin{align*}
    (1-(p_1+\epsilon))(1-(p_2-\epsilon))\prod_{i=3}^{t}(1-p_i)-\prod_{i=1}^t(1-p_i)
    =\epsilon(p_2-p_1-\epsilon)\prod_{i=3}^{t}(1-p_i)<0.
  \end{align*}
  This contradicts the fact the quantity is minimized.  Hence, there is at
  most one $p_i$ which is strictly between $0$ and $\pmax$.

  The last inequality holds since $1-b\geq (1-a)^{b/a}$ for any $0\leq b\leq
  a\leq 1$.
\end{pf}

Let $\partial_{\hE}(e)$ be the set of edges in $\hE$ incident on either endpoint of $e$ excluding $e$ itself.

\begin{lem}
  \label{lem:prob-probed-random}
  Let $e$ be an edge in $\hE$ and let $\sigma$ be a random edge
  ordering. Let $\pmax=\max_{f\in \hE}p_{f}$. Assume
  that $\sum_{f\in \partial_{\hE} (e)}p_{f} = r$.  Then,
  \[ \E_\sigma\Bigr[\, \prod_{f \in B(e, \sigma)}
    (1-p_f)\, \mid \, \hE \, \Bigr] \geq
  \int_{0}^{1} \eta(xr, xp_{\max}) \d x. \]
\end{lem}
\begin{pf}
We claim that the expectation can be written in the following continuous form:
\begin{align*}
\E_\sigma\Bigr[\,  \prod_{f \in B(e, \sigma)} (1-p_f)\, \mid \,
\widehat{E} \, \Bigr] =\int_{0}^{1} \prod_{f\in \partial_{\hE} (e)} (1-xp_f)
\d x.
\end{align*}
The lemma easily follows from this and Lemma~\ref{lem:eta}.

To see the claim, we consider the following random experiment: For
each edge $f\in \partial(e)$, we pick uniformly at random a real
number $a_f$ in $[0,1]$. The edges are then sorted according to these
numbers.  It is not difficult to see that the experiment produces
uniformly random orderings. For each edge $f$, let the random variable
$A_f=1-p_f$ if $f\in B(e,\sigma)$ and $A_f=1$ otherwise.  Hence, we
have
\begin{align*}
 \E_\sigma\Bigl[\, \prod_{f \in B(e, \sigma)} (1-p_f)\,\mid\,\widehat{E}\, \Bigr]
& =  \int_{0}^{1}\E\Bigl[\, \prod_{f \in
\partial_{\hE}(e)} A_f\,\mid\, a_e=x\,\Bigr]\d x \\
& = \int_{0}^{1}\prod_{f \in \partial_{\hE}(e)}\E\Bigl[\,  A_f\,\mid\,
a_e=x \, \Bigr]\d x \\
&=  \int_{0}^{1}\prod_{f
  \in \partial_{\hE}(e)}\bigl(x(1-p_f)+(1-x)\bigr)\d x \\
& =\int_{0}^{1}\prod_{f \in
\partial_{\hE}(e)}(1-xp_f)\d x
\end{align*}
The second equality holds since the $A_f$ variables, conditional on $a_e=x$, are independent.
\end{pf}

\begin{lem}
  \label{lem:prob-probing}
  Let $\rho(r,\pmax)=\int_{0}^{1} \eta(xr, xp_{\max}) \d x$. For any $r,\pmax>0$, we have
  \begin{enumerate}
    \item $\rho\,(r,\pmax)$ is convex and decreasing on $r$.
    \item $\rho\,(r,\pmax)\geq {1\over r+p_{\max}}\cdot\Bigl(1-(1-p_{\max})^{1+\frac{r}{p_{\max}}}\Bigr) > {1\over r+p_{\max}}\cdot\Bigl(1-e^{-r}\Bigr)$
    \end{enumerate}
\end{lem}

\begin{pf}
  To see the first part, let us consider the function values on discrete
  points $r=\pmax, 2\pmax, \ldots$.  Let $F(x)=\frac{1}{x}(1-c^x)$
  where $c=1-\pmax$.  From the above derivation, we can easily get
  that for integral $t$,
  \begin{align*}
    \rho\,(t\pmax,\pmax) = \int_{0}^{1} (1-xp_{\max})^{t} \d x =
    \frac{1}{\pmax(t +1)} \left(1-c^{t +1}\right) = {1\over \pmax}F(t+1).
  \end{align*}
  The function $F(x)$ is a convex function for any $0<c<1$.  Indeed, it is
  not hard to prove that $\frac{\d^2}{\d x^2} F(x)= \frac{2}{x^3}+ c^x\left(
    -\frac{2}{x^3}+\frac{2\ln a}{x^2}-\frac{\ln^2 a}{x}\right)>0$ for any
  $0<c<1$.  However, $\rho\,(t\pmax,\pmax)$ only coincides with ${1\over
    \pmax}F(t+1)$ at integral values of $t$. Now, let us consider the value of
  $\rho(r,\pmax)$ for $\gamma \pmax< r <(\gamma+1)
  \pmax$:
  \begin{align}
    \label{eq-rho}
    \rho\,(r,\pmax) & = \int_{0}^{1} (1-x \pmax)^{\gamma} \Bigl(1-x(r-\gamma
    \pmax)\Bigr) \d x
\end{align}

\noindent
The key observation is that for fixed values of $\pmax$ and $\gamma$
the right hand side of \eqref{eq-rho} is a just linear function of
$r$. The dependency of $\rho$ in terms of $r$ then becomes clear: it
is a piecewise linear function that takes the value $F(t+1)$ at
abscissa points $t\pmax$ for $t\in \mathbb{Z}_0$. Therefore, $\rho$ is
a convex decreasing function of $r$.

The second part follows easily from Lemma~\ref{lem:eta}:

\begin{align*}
  \rho\,(r,\pmax) & = \int_{0}^{1}\eta(xr,xp_{\max}) \d x
  \geq \int_{0}^{1} (1-xp_{\max})^{r/p_{\max}}  \d x \\
  & = {1\over r+p_{\max}}\cdot\Bigl(1-(1-p_{\max})^{1+\frac{r}{p_{\max}}}\Bigr)
  \geq{1\over r+p_{\max}}\cdot\Bigl(1-e^{-r}\Bigr)
\end{align*}

\end{pf}

\begin{lem}
  \label{lem:decoupling}
  Let $e =(u,v) \in \hE$. Let $\pi$ be a random matching ordering and
  $\sigma$ be a random edge ordering of the edges adjacent to $u$ and $v$. Then
  \[ \E_\pi \Bigr[ \prod_{f \in B(e,\pi)} (1-p_f) \,\mid\,
    \widehat{E} \, \Bigr] \geq \E_\sigma
  \Bigr[ \prod_{f \in B(e,\sigma)} (1-p_f)\, \mid\, \widehat{E}\, \Bigr]. \]

\end{lem}

\begin{pf}
  We can think of $\pi$ as a permutation of bundles of edges: For each
  matching, if there are two edges incident on $e$, we bundle the edges
  together; if there is a single edge incident on $e$ this edge is in a
  singleton bundle by itself.  The random edge ordering $\sigma$ can be
  thought as having all edges incident on $e$ in singleton bundles.

  Consider the same random experiment as in Lemma~\ref{lem:prob-probed-random}
  except that we only pick one random number for each bundle.
  Let $G(e)$ be the set of all bundles incident on $e$. Using the same argument,
  we have
  \[ \E_\pi \Bigr[ \prod_{f \in B(e,\pi)} (1-p_f) \, \mid\,
    \widehat{E}\,\Bigr] =\int_{0}^{1} \prod_{g\in G(e)}
  \Bigl(x\cdot\prod_{f\in g}(1-p_f)+(1-x)\Bigr) \d x. \]
  But for any bundle $g\in G(e)$ and $0\leq x\leq 1$, we claim that
  \[x\cdot\prod_{f\in g}(1-p_f)+(1-x)\geq \prod_{f\in g}(1-xp_f).\]
  For singleton bundles we actually have equality. For a bundle
  $g=\{f_1,f_2\}$, we have $x(1-p_{f_1})(1-p_{f_2})+(1-x) =
  1-xp_{f_1}-xp_{f_2}+xp_{f_1}p_{f_2} \geq
  1-xp_{f_1}-xp_{f_2}+x^2p_{f_1}p_{f_2} =
  (1-xp_{f_1})(1-xp_{f_2})$. This completes the proof.
\end{pf}

As we shall see shortly, if $\sum_{f \in \partial_{\hE}(e)} p_e$ is
small then the probability that $e$ is not blocked is large. Because
of the marginal distribution property of the GKSP rounding procedure,
we can argue that this quantity is small \emph{in expectation} since
$\sum_{f \in \partial(e)} p_e y_e \leq 2$ due to the fact that $y$ is
a feasible solution to \eqref{LP:stoc-matching}. This, however, is not
enough; in fact, for our analysis to go through, we need a slightly
stronger property.

\begin{lem}
  \label{lem:exp-success-probe}
  For every edge $e$,
  \[ \E \Big[ \sum_{f \in \partial_{\hE}(e) } p_f  \ \mid \
  e \in \hE\,\Big] \leq \sum_{f \in \partial(e) } p_f \, y_f. \]
\end{lem}

\begin{pf} Let $u$ be an endpoint of $e$.
  \begin{align*}
     \E \Big[ \sum_{f \in \partial_{\hE}(u) - e} p_f \ \mid \
  e \in \hE \, \Big]
  & = \sum_{f \in \partial(u) - e} \Pr[\,\widehat{y}_f =1
  \,\mid\,\widehat{y}_e = 1]\, \cdot p_f,  \\
  &  \leq \sum_{f \in \partial(u) - e} \Pr[\,\widehat{y}_f =1\,] \cdot
  p_f, & \text{[by Theorem~\ref{thm:dep-rounding} P3]} \\
  &  = \sum_{f \in \partial(u) - e} y_f \, p_f. & \text{[by
    Theorem~\ref{thm:dep-rounding} P1]}.
  \end{align*}

  The same bound holds for the other endpoint of $e$. Adding the two
  inequalities we get the lemma.
\end{pf}

Everything is in place to derive a bound the expected weight of the
matching found by our algorithm.

\begin{thm}
  \label{thm:non-uniform-bipartite}
  If $G$ is bipartite then {\sc round-color-probe} is a
  $1/\rho(2,\pmax)$ approximation under the edge- and
  matching-probing model, where $\rho$ is defined in
  Lemma~\ref{lem:prob-probing}. The worst ratio is attained at $\pmax
  =1$, where it is $3$. The ratio tends to $\frac2{1-e^{-2}}$ as
  $\pmax$ tends to 0.
\end{thm}

\begin{pf}
  Recall that the optimal value of \eqref{LP:stoc-matching} is exactly
  $\sum_{e \in E} w_e y_e x_e.$ The expected weight
  of the matching found by the algorithm is
  \begin{align*}
    \E\,[\,\algo\,] & = \sum_{e \in E} w_e\, p_e\, \Pr [\, e \in \hE\, ]
    \cdot \Pr\,[\, e \text{ was not
      blocked } \, \mid\, e \in \widehat{E} \,] \\
    & = \sum_{e \in E} w_e\, p_e \, y_e \cdot \Pr\,[\, e \text{ was
      not blocked }  \, \mid\, e \in \widehat{E} \, ] & \text{[by
      Theorem~\ref{thm:dep-rounding} P1]}\\
    &\geq \sum_{e \in E} w_e\, p_e \, y_e \cdot \E_\pi\,\Bigl[\,
    \prod_{f \in B(e, \pi)}
    (1-p_f)\, \mid \, e \in \hE \, \Bigr] & \text{[by \eqref{eq:prob-probed}]}\\
    & \geq \sum_{e \in E} w_e\, p_e \, y_e \cdot \E_\sigma\,\Bigl[\,
    \prod_{f \in B(e, \sigma)}
    (1-p_f)\, \mid \, e \in \hE \, \Bigr] & \text{[by Lemma~\ref{lem:decoupling}]} \\
    & \geq \sum_{e \in E} w_e\, p_e \, y_e \cdot \E \Bigr[\, \rho
    \Bigr(\sum_{f \in \partial_{\widehat{E}}(e)} p_f, \pmax \Bigr) \,
    \mid \, e \in \hE \, \Bigr] & \text{[by
      Lemma~\ref{lem:prob-probed-random}]}\\
    & \geq \sum_{e \in E} w_e\, p_e \, y_e \cdot \rho
    \Bigl(\E\,\Bigl[\sum_{f \in \partial_{\hE}(e)} p_f \, \mid \, e \in
    \widehat{E}\,\Bigl],
    \pmax\Bigl) & \text{[by Jensen's inequality]} \\
    & \geq \sum_{e \in E} w_e\, p_e \, y_e \cdot \rho
    \Bigl(\sum_{f \in \partial(e)} y_f\, p_f,
    \pmax\Bigl) & \text{[by Lemma~\ref{lem:exp-success-probe}]} \\
    & \geq \sum_{e \in E} w_e\, p_e \, y_e \cdot \rho (2, \pmax) &
    \text{[$y$ is feasible for \eqref{LP:stoc-matching}]}.
  \end{align*}
  Notice that we are able to use Jensen's inequality because, as shown
  in Lemma~\ref{lem:prob-probing}, $\rho(r,\pmax)$ is a convex
  and decreasing function of $r$. The last inequality also uses the
  fact that $\rho$ is decreasing.

It can be checked directly (using the first inequality in Lemma~\ref{lem:prob-probing}(2)) that $\rho(2,p_{max})$ is
maximized at $p_{max}=1$ where it is 3. Moreover $\rho(2,p_{max})\rightarrow \left(1-e^{-2}\right)/2$ as $p_{max}$
tends to $0$.
\end{pf}

\subsection{Weighted Stochastic Matching: General Graphs Redux} \label{subsec:wt-gen}

We present an alternative algorithm for weighted stochastic matching in general graphs that builds on the algorithm for
the bipartite case. The basic idea is to solve \eqref{LP:stoc-matching}, randomly partition the vertices of $G$ into
two sets $A$ and $B$, and then run {\sc round-color-probe} on the bipartite graph induced by $(A,B)$. For the analysis
to go through, it is crucial that we use the already computed fractional solution instead of solving again
\eqref{LP:stoc-matching} for the new bipartite graph in the call to {\sc round-color-probe}.

\begin{shadebox}
  \begin{OneLiners}
    \item[1.] $(x,y)$ \assign\ optimal solution to \eqref{LP:stoc-matching}
    \item[2.] randomly partition vertices into $A$ and $B$
    \item[3.] run {\sc round-color-probe} on the bipartite graph and
    the fractional solution induced by $(A,B)$
  \end{OneLiners}
\end{shadebox}

\begin{thm}
  \label{thm:non-uniform-general}
  For general graphs there is a $2/\rho(1,\pmax)$ approximation under
  the edge- and matching-probing model, where $\rho$ is defined in
  Lemma~\ref{lem:prob-probing}. The worst ratio is attained at $\pmax
  =1$, where it is $4$. The ratio tends to $\frac2{1-e^{-1}}$ as
  $\pmax$ tends to 0.
\end{thm}

\begin{pf}
  The analysis is very similar to the bipartite case. Essentially,
  conditional on a particular outcome for the partition $(A,B)$, all
  the lemmas derived in the previous section hold. In other words, the
  same derivation done in the proof of
  Theorem~\ref{thm:non-uniform-general} yields:
  \begin{equation*}
    \E [\, \algo\, \mid \, (A,B) \,] \geq \sum_{e \in (A,B)} w_e p_e
    y_e \cdot \rho \Bigr(\sum_{f \in \partial_{A,B}(e)} p_f\, y_f, \pmax\Bigr),
  \end{equation*}
  where $\partial_{A,B}(e) = \partial(e) \cap (A,B)$.

  Hence, the expectation of algorithm's performance is:
  \begin{align*}
    \E [\, \algo\, ] & \geq \sum_{e \in E} w_e \, p_e\,
    y_e \, \Pr [ e \in (A,B) ] \cdot \E \Bigr[ \rho \Bigr(\sum_{f
      \in \partial_{A,B}(e)} p_f\, y_f, \, \pmax\Bigr) \, \mid \, e \in
    (A,B) \, \Bigr], \\
    & \geq  \sum_{e \in E} w_e \, p_e \,
    y_e \, \frac{1}{2} \cdot \rho \Bigr(\E \Bigr[\sum_{f
      \in \partial_{A,B}(e)} p_f\, y_f \, \mid \, e \in (A,B)
    \Bigr], \, \pmax\Bigr), \\
    & \geq  \sum_{e \in E} w_e \, p_e \,
    y_e \, \frac{1}{2} \cdot \rho \Bigr(\sum_{f
      \in \partial(e)} \frac{p_f \, y_f}{2}, \, \pmax\Bigr), \\
    & \geq  \sum_{e \in E} w_e \, p_e \,
    y_e \, \frac{1}{2} \cdot \rho \big(1, \,\pmax\big),
  \end{align*}
  where the second inequality follows from Jensen's inequality and the
  fact that $\rho(r,\pmax)$ is a convex decreasing function of
  $r$. Finally, noting that $\sum_{e \in E} w_e \, p_e\, y_e$ is a
  lower bound on the value of the optimal strategy, the theorem
  follows.
\end{pf}

 \subsection{Unweighted Stochastic Matching}
\label{subsec:unwt}

In this subsection, we consider the unweighted stochastic matching problem, and show that our algorithm from
\S\ref{subsec:wt-gen} can be combined with the natural greedy algorithm~\cite{c-etal} to obtain a better approximation
guarantee than either algorithm can achieve on their own. Basically, our algorithm attains its worst ratio when $\pmax$
is large and greedy attains its worst ratio when $\pmax$ is small. Therefore, we can combine the two algorithms as
follows: We probe edges using the greedy heuristic until the maximum edge probability in the remaining graph is less
than a critical value $p_c$, at which point we switch to our algorithm from \S\ref{subsec:wt-gen}.  We denote by
$\algo$ this combined algorithm and by $\opt$ the optimal probing strategy.

\begin{lem}
  \label{lem:combine-approx}
  Suppose that we use the greedy rule until all remaining edges have
  probability less than $p_c$, at which point we switch to an
  algorithm with approximation ratio $\gamma(p_c)$.  Then the
  approximation ratio of the overall scheme is $\alpha(p_{c})=
  \max\set{4-p_c, \gamma(p_c)}$.
\end{lem}

\begin{pf}
  First, let us review some facts from the work of
  Chen~\etal~\cite{c-etal}.  Let $(G,t)$ be an instance of the
  edge-probing model.  Suppose $e=(u,v)$ is the edge with the largest
  probability.  Denote by $(G_L,t_L)$ and $(G_R,t_R)$ the instances
  resulted from the success and failure for the probe to $e$,
  respectively.  In other words, $G_L=G\setminus \{u,v\}, t_L=t$ and
  $G_R=G\setminus \{e\}, t_R(u)=t(u)-1, t_R(v)=t(v)-1,
  t_R(w)=t(w)\,\,\forall w\ne u,v$.  Denote the expected value
  generated by algorithm $\algo$ on instance $(G,t)$ by
  $\E_{\algo}(G,t)$. Suppose $p_{max} > p_c$. It is easy to see that,
  for any $\algo$ that first probes $e$,
  \begin{align}
    \label{eq:exp} \E_{\algo}(G,t)=p_e+p_e\E_{\algo}(G_L, t_L)+(1-p_e)\E_{\algo}(G_R, t_R).
  \end{align}
  Moreover, Chen \etal\ showed that
  \begin{align}
    \label{eq:opt} \E_{\opt}(G,t)\leq p_e(4-p_e)+p_e\E_{\opt}(G_L, t_L)+(1-p_e)\E_{\opt}(G_R, t_R).
  \end{align}
  Now, we prove the theorem by induction on the size (the number of
  vertices and edges) of the instance.  The base cases are all
  instances in which the maximum probability is at most $p_c$.  Then $
  \E_{\algo}(G,t)\leq \gamma(p_c)\E_{\opt}(G,t)\leq
  \alpha(p_c)\E_{\opt}(G,t) $ for any base instance.  The inductive
  step only concerns instances where greedy is used.  Thus, following
  from \eqref{eq:exp} and \eqref{eq:opt} and the inductive hypothesis,
  we get
  \begin{align*}
    \E_{\opt}(G,t) & \leq   p_{e}\alpha(p_{c})+p_{e} \alpha(p_{c}) \E_{\algo}(G_{L}, t_{L})
    +(1-p_{e})\alpha(p_{c})\E_{\algo}(G_{R},t_{R})\\ & \leq  \alpha(p_{c}) \E_{\algo}(G,t).
  \end{align*}
  This completes the inductive proof.
\end{pf}

We are ready to derive the improved bounds for
unweighted stochastic matching.

\medskip

\begin{proofof}{Theorem~\ref{th:main1.5}}
  The stated approximation guarantee can be obtained by setting the
  cut-off point to $p_c = 0.541$ and then using
  Lemma~\ref{lem:combine-approx} in combination with
  Theorem~\ref{thm:non-uniform-general} for bounding the performance
  of the second algorithm at $\pmax= p_c$.
\end{proofof}

\medskip

We remark that the approximation ratio of the algorithm in Section~\ref{subsec:wt-match} does not depend on $\pmax$,
thus we can not combine that algorithm with the greedy algorithm to get a better bound. Furthermore, the result of this
subsection only holds for the unweighted version of the problem since greedy has an unbounded approximation ratio in
the weighted case.


\section{Stochastic \emph{Online} Matching (Revisited)}
\label{sec:online}

\def\ds{\ensuremath{\mathcal{D}}\xspace}
\def\a{\ensuremath{\mathcal{A}}\xspace}
\def\b{\ensuremath{\mathbf{b}}}
\def\lp{\ensuremath{{\sf LP}}\xspace}
\def\sop{\ensuremath{{\sf Stoc_On_Match}}\xspace}

As mentioned in the introduction, the stochastic online matching problem
is best imagined as selling a finite set of goods to buyers that arrive
over time. The input to the problem consists of a bipartite graph $G=(A,
B,A\times B)$, where $A$ is the set of {\em items} that the seller has to
offer, with exactly one copy of each item, and $B$ is a set of {\em
  buyer types/profiles}. For each buyer type $b\in B$ and item $a\in A$,
$p_{ab}$ denotes the probability that a buyer of type $b$ will like
item $a$, and $w_{ab}$ denotes the revenue obtained if item $a$ is sold
to a buyer of type $b$. Each buyer of type $b\in B$ also has a patience
parameter $t_b \in \mathbb{Z}_+$. There are $n$ buyers arriving online,
with $e_b \in \mathbb{Z}$ denoting the expected number of buyers of type
$b$, with $\sum e_b = n$. Let \ds denote the induced probability
distribution on $B$ by defining $\Pr_\ds[b]=e_b/n$. All the above
information is given as input.

The stochastic online model is the following: At each point in time, a
buyer arrives, where her type $\mathbf{b} \in_{\ds} B$ is an i.i.d.\
draw from \ds. The algorithm now shows her \emph{up to $t_\b$ distinct
  items one-by-one}: the buyer likes each item $a\in A$ shown to her
independently with probability $p_{ab}$. The buyer purchases the first item that she is {\em offered and likes}; if
she buys item $a$, the revenue accrued is $w_{ab}$. If she does not like any of the items shown, she leaves without
buying.  The objective is to maximize the expected revenue.

We get the stochastic online matching problem of Feldman~\etal~\cite{FMMM} if we have $w_{ab} = p_{ab} \in \{0,1\}$, in
which case we need only consider $t_b = 1$. Their focus was on beating the $1-1/e$-competitiveness known for worst-case
models~\cite{KarpVV90,KalyanasundaramP93,MSVV05,BirnbaumM08,GoelM08}; they gave a $0.67$-competitive algorithm that
works for the unweighted case with high probability. On the other hand, our results are for the weighted case (with
preference-uncertainty and timeouts), but only in expectation. Furthermore, in our extension, due to the presence of
timeouts (see \S\ref{subsec:prelim}), any algorithm that provides a guarantee whp must necessarily have a high
competitive ratio.

{\em By making copies of buyer types, we may assume that $e_b =1$ for all $b\in B$, and \ds is uniform over $B$.}  For
a particular run of the algorithm, let $\hat B$ denote the actual set of buyers that arrive during that run.  Let $\hat
G = (A, \hat B, A \times \hat B)$, where for each $a\in A$ and $\hat b \in \hat B$ (and suppose its type is some $b\in
B$), the probability associated with edge $(a,\hat b)$ is $p_{ab}$ and its weight is $w_{ab}$. Moreover, for each $\hat
b\in \hat B$ (with type, say, $b\in B$), set its patience parameter to $t_{\hat b} = t_b$. We will call this the
\emph{instance graph}; the algorithm sees the vertices of $\hat B$ in random order, and has to adaptively find a large
matching in $\hat G$.

It now seems reasonable that the algorithm of \S\ref{subsec:wt-match} should work here. But the algorithm does
not know $\hat G$ (the actual instantiation of the buyers) up front, it only knows $G$, and hence some more work is
required to obtain an algorithm.  Further, as was mentioned in the preliminaries, we use \opt to denote the optimal
adaptive strategy (instead of the optimal offline matching in $\hat G$ as was done in \cite{FMMM}), and compare our
algorithm's performance with this \opt.


\medskip
\noindent {\bf The Linear Program.} For a graph $H = (A, C, A \times
C)$ with each edge $(a,c)$ having a probability $p_{ac}$ and weight
$w_{ac}$, and vertices in $C$ having patience parameters $t_j$,
consider the LP$(H)$:
\begin{linearprogram}
  {
    \label{LP:stoc-online}
    \maximize{\displaystyle \sum_{a\in A,\,c\in C} \,\, w_{ac}\cdot x_{ac}}
  }
  \sum_{c\in C} x_{ac} &\le 1& &\forall a\in A\label{eq:os-lp2}\\
  \sum_{a\in A} x_{ac} &\le 1& &\forall c\in C\label{eq:os-lp3}\\
  \sum_{a\in A} y_{ac} &\le t_c & &\forall c\in C\label{eq:os-lp4}\\
  x_{ac}&=p_{ac}\cdot y_{ac}& &\forall a\in A,\,c\in C \label{eq:os-lp5}\\
  y_{ac}&\in [0,1]& &\forall a\in A,\,c\in C \label{eq:os-lp6}
\end{linearprogram}

Note that this LP is very similar to the one in \S\ref{sec:stoc-match}, but the vertices on the left do not have
timeout values. Let $\lp(H)$ denote the optimal value of this LP. \\

\noindent {\bf The algorithm:}
\begin{shadebox}
  \begin{OneLiners}
  \item[1.] Before any buyers arrive, solve the LP on the expected
    graph $G$ to get values $y^*$.
  \item[2.] When any buyer $\hat b$ (of type $b$) arrives online:
    \begin{OneLiners}
    \item[a.] If $\hat b$ is the first buyer of type $b$, consider the
      items $a\in A$ in u.a.r. order. One by one, offer each item $a$
      (that is still unsold) to $\hat b$ independently with
      probability $y^*_{ab}/\alpha$; stop if either $t_b $ offers are
      made or $\hat b$ purchases any item.
    \item[b.] If $\hat b$ is not the first arrival of type $b$, do not offer
      any items to $\hat b$.
    \end{OneLiners}
  \end{OneLiners}
\end{shadebox}
In the following, we prove that our algorithm achieves a constant approximation to the stochastic online matching
problem. The first lemma show that the expected value obtained by the best online adaptive algorithm is bounded above
by $\E[\lp(\hat G)]$.

\begin{lem}
  \label{lem:stoc-on0}
  The optimal value $\opt$ of the given instance is at most $\E[\lp(\hat
  G)]$, where the expectation is over the random draws to create $\hat
  G$.
\end{lem}
\begin{pf}
  Consider an algorithm that is allowed to see the instantiation $\hat
  B$ of the buyers before deciding on the selling strategy---the
  expected revenue of the best such algorithm is clearly an upper bound
  on \opt. Given any instantiation $\hat B$, the expected revenue of the
  optimal selling strategy is at most $\lp(\hat G)$ (see e.g.
  Claim~\ref{clm:adaptive}). The claim follows by taking an expectation
  over $\hat B$.
\end{pf}

The proof of the next lemma is similar to the analysis of
Theorem~\ref{th:main1} for weighted stochastic matching.

\begin{lem}\label{lem:stoc-on1}
  Our expected revenue is at least $ \left(1-\frac1e\right)\, \frac1\alpha \,
  \left(1-\frac1\alpha - \frac2{3\alpha^2}\right)\cdot \lp(G)$.
\end{lem}

\begin{pf}
  For any buyer-type $b\in B$, in this proof, $\hat b$ refers to the
  first type-$b$ buyer (if any). For each $b\in B$, let r.v.
  $T_b\in[n]\cup\{\infty\}$ denote the earliest arrival time of a
  type-$b$ buyer; if there is no type-$b$ arrival then $T_b=\infty$.
  Note that our algorithm obtains positive revenue only for buyers
  $\{\hat b\mid b\in B,\, T_b<\infty\}$; let $R_b$ denote the revenue
  obtained from buyer $\hat b$ (if any). The expected revenue of the
  algorithm is $\E[\sum_{b\in B} R_b]$. We now estimate $\E[R_b]$ for a
  fixed $b\in B$.

  Let $\a_b\equiv (T_b<\infty)$ denote the event that there is some
  type-$b$ arrival in the instantiation $\hat B$. Since each arrival is
  i.i.d. from the uniform distribution over $B$, $\Pr[\a_b]=
  1-(1-1/n)^n\ge 1-\frac1e$. In the following, we condition on $\a_b$
  and bound $\E[R_b \mid \a_b]$. Hence we assume that buyer $\hat b$
  exists.

  For any vertex $a\in A$, let $M_a$ denote the indicator r.v. that $a$
  is already matched before time $T_b$; and $O_a$ (resp. $M'_a$) the
  indicator r.v. that $\hat b$ is timed-out (resp. already matched) when
  item $a$ is considered for offering to $\hat b$. Now,
  \begin{align}
    \Pr[\, \mbox{item $a$ offered to }\hat b\mid \a_b\,] &=
    (1-\Pr[\,M_a\cup M'_a\cup
    O_a \mid \a_b\,])\cdot \frac{y_{ab}}{\alpha}\notag \\
    &\ge (1- \Pr[\, M_a\mid \a_b\,]- \Pr[\, M'_a\cup O_a \mid \a_b\,])\cdot
    \frac{y_{ab}}{\alpha} \label{eq:stoc-on1}
  \end{align}

  \begin{cl}\label{cl:stoc-on1}
    For any $a\in A$ and $b\in B$, $\Pr[\,M_a\mid \a_b\,]\le
    \frac1{2\alpha}$.
  \end{cl}
  \begin{pf}
    For any $v\in B\setminus \{b\}$, let $I_b^v$ denote the indicator
    r.v. for the event $T_v<T_b$. We have:
    \begin{align}
      \Pr[\,M_a\mid \a_b\,] &= \sum_{v\in B\setminus \{b\}}
      \Pr[\,\mbox{type-$v$ buyer is matched to $a$ before time $T_b$} \mid
      \a_b\,]\\
      & = \sum_{v\in B\setminus \{b\}} \Pr[\,I_b^v \mid \a_b\, ] \cdot
      \Pr[\, \hat v \mbox{ matched to }a \mid I_b^v,\,\a_b \,]\\
      & \le  \sum_{v\in B\setminus \{b\}} \Pr[\,I_b^v\mid \a_b\, ] \cdot
      \frac{x_{av}}\alpha \quad \le \quad \frac12 \sum_{v\in B\setminus
        \{b\}} \frac{x_{av}}\alpha \quad \le \quad \frac1{2\alpha},
    \end{align}
    where the first inequality follows from the fact that even after the
    algorithm has considered an edge $(a,v)$, the probability of
    matching $(a,v)$ is $\frac{y_{av}}\alpha\cdot p_{av}$, the last
    inequality uses LP-constraint~\eqref{eq:os-lp2} for graph $G$, and
    the second last inequality uses $\Pr[\,I_b^v\mid \a_b\,]\le \frac12$
    (for $v\in B\setminus \{b\}$), which we show next.

    Note that event $I_b^v\wedge \a_b$ corresponds to $(T_v < T_b
    <\infty)$; and event $\a_b$ contains both $(T_v < T_b <\infty)$ and
    $(T_b < T_v <\infty)$. By symmetry, $\Pr[\,T_v < T_b <\infty\,] =
    \Pr[\,T_b < T_v <\infty\,]$, which implies:
    $$\Pr[\,I_b^v\mid \a_b\,]  = \frac{\Pr[\,T_v < T_b <\infty\,]} {\Pr[\a_b]  }
    \le \frac{\Pr[\,T_v < T_b <\infty\,]} {\Pr[\, (T_v < T_b
      <\infty) \vee (T_b < T_v <\infty)\,]} =\frac12.$$ This completes the
    proof of Claim~\ref{cl:stoc-on1}.
  \end{pf}

  \begin{cl}
    \label{cl:stoc-on2}
    For any $a\in A$ and $b\in B$, $\Pr[\,M'_a\cup O_a \mid \a_b\,]\le
    \frac1{2\alpha}+\frac2{3\alpha^2}$.
  \end{cl}
  \begin{pf}
    This is a direct application of Lemmas~\ref{lem:match-prob}
    and~\ref{lem:high-TO-prob}, since items offered to $\hat b$ are
    considered in u.a.r. order. As in \S\ref{subsec:wt-match}, there
    are two cases:
    \begin{itemize}
    \item Suppose $t_b =1$. Here we have $M'_a\sse O_a$, so
      $\Pr[\,M'_a\cup O_a \mid \a_b\,]=\Pr[\,O_a \mid \a_b\,]\le
      \frac1{2\alpha}$, by the proof of Lemma~\ref{lem:match-prob} using
      LP-constraint~\eqref{eq:os-lp4}.

    \item Suppose $t_b \ge 2$. Using the proof of
      Lemma~\ref{lem:high-TO-prob} and LP-constraint~\eqref{eq:os-lp4},
      we have $\Pr[\,O_a\mid \a_b\,]\le \frac2{3\alpha^2}$. Again by the
      proof of Lemma~\ref{lem:match-prob} and
      LP-constraint~\eqref{eq:os-lp3}, $\Pr[\,M'_a\mid \a_b\,]\le
      \frac1{2\alpha}$.
    \end{itemize}
    In both cases above, the statement in Claim~\ref{cl:stoc-on2} holds.
  \end{pf}

Now applying Claims~\ref{cl:stoc-on1} and~\ref{cl:stoc-on2}
  to~\eqref{eq:stoc-on1}, we obtain:
  $$\Pr[\,\mbox{item $a$ offered to }\hat b\mid \a_b\,] \ge \frac1\alpha
  \left(1-\frac1\alpha - \frac2{3\alpha^2}\right)\cdot y_{ab}.$$ This
  implies:
  \begin{align*}
  \E[R_b\mid \a_b] & = \sum_{a\in A} w_{ab}\cdot p_{ab}\cdot
  \Pr[\,\mbox{item $a$ offered to }\hat b\mid \a_b\,] \\
   & \ge  \frac1\alpha
  \left(1-\frac1\alpha - \frac2{3\alpha^2}\right)\sum_{a\in A}
  w_{ab}\cdot x_{ab}.
  \end{align*}
  Since $\Pr[\,\a_b\,]\ge 1-\frac1e$, we also have
  $\E[R_b] \ge (1-\frac1e)\, \frac1\alpha \, \left(1-\frac1\alpha -
    \frac2{3\alpha^2}\right)\sum_{a\in A} w_{ab}\cdot x_{ab}$.

  Finally, the expected revenue obtained by the algorithm is:
  $$\sum_{b\in B} \E[R_b]\ge
  \left(1-\frac1e\right)\, \frac1\alpha \, \left(1-\frac1\alpha -
    \frac2{3\alpha^2}\right)\cdot \lp(G).$$ This proves
  Lemma~\ref{lem:stoc-on1}.
\end{pf}

Note that we have shown that $\E[\lp(\hat G)]$ is an upper bound on
\opt, and that we can get a constant fraction of $\lp(G)$. The final
lemma relates these two, namely the LP-value of the expected graph $G$
(computed in Step~1) to the expected LP-value of the instantiation $\hat
G$; the proof uses a simple but subtle duality-based argument.
\begin{lem}\label{lem:stoc-on2}
  $\lp(G)\ge \E[\lp(\hat G)]$.
\end{lem}
\begin{pf}
  Consider the dual of the linear
  program~\eqref{LP:stoc-online}.
  \begin{align}
    &\min  \sum_{a\in A} \alpha_a +\sum_{c\in C} \left( \alpha_c +
      t_c \cdot \beta_c\right) + \sum_{a\in A,\,c\in C} z_{ac} &\\
    &z_{ac} + p_{ac} \cdot (\alpha_a +\alpha_c) + \beta_c \ge
    w_{ac}\cdot p_{ac} & \forall a\in A,\,c\in C   \label{eq:os-dp2}\\
    & \alpha,\beta,z \ge 0&\label{eq:os-dp3}
  \end{align}
  Let $(\alpha,\beta,z)$ denote the optimal dual solution corresponding
  to graph $G$; note that its objective value equals $\lp(G)$ by strong
  duality. For any instantiation $\hat G$, define dual solution $(\hat
  \alpha,\hat \beta, \hat z)$ as follows:
  \begin{OneLiners}
  \item For all $a\in A$, $\hat \alpha_a=\alpha_a$.
  \item For each $c \in \hat B$ (of type $b$), $\hat \alpha_c = \alpha_b$ and $\hat{\beta}_c=\beta_b$.
  \item For each $a\in A$ and $c \in \hat B$ (of type $b$), $\hat z_{ac}
    = z_{ab}$.
  \end{OneLiners}
  Note that $(\hat \alpha,\hat \beta, \hat z)$ is a feasible dual
  solution corresponding to the LP on $\hat G$: there is constraint for
  each $a\in A$ and $c\in \hat B$, which reduces to a constraint for
  $(\alpha,\beta,z)$ in the dual corresponding to $G$.  By weak duality,
  the objective value for $(\hat \alpha,\hat \beta, \hat z)$ is an
  upper-bound on $\lp(\hat G)$. For each $b\in B$, let $N_b$ denote the
  number of type $b$ buyers in the instantiation $\hat B$; note that
  $\E[N_b]=1$ by definition of distribution \ds. Then the dual objective
  for $(\hat \alpha,\hat \beta, \hat z)$ satisfies:
  \[  \sum_{a\in A} \alpha_a +\sum_{b\in B} N_b \cdot \left( \alpha_b
    + t_b \cdot \beta_b\right) + \sum_{a\in A,\,b\in B} N_b\cdot z_{ab}
  \quad \ge \quad \lp(\hat G). \] Taking an expectation over $\hat B$,
  we obtain:
  \begin{align*}
    \E[\lp(\hat G)] & \le   \sum_{a\in A} \alpha_a + \sum_{b\in B}
    \E[N_b] \cdot \left(  \alpha_b + t_b \cdot \beta_b + \sum_{a\in A}
      z_{ab}\right)\\
    &=   \sum_{a\in A} \alpha_a +\sum_{b\in B} \left( \alpha_b +
      t_b \cdot \beta_b\right) + \sum_{a\in A,\,b\in B} z_{ab}
    = \lp(G).
  \end{align*}
  This proves the lemma.
\end{pf}

\noindent Applying Lemmas~\ref{lem:stoc-on0},~\ref{lem:stoc-on1}
and~\ref{lem:stoc-on2}, and setting $\alpha=\frac2{\sqrt{3}-1}$,
completes Theorem~\ref{th:main3}'s proof.

\section{Stochastic $k$-Set Packing}
\label{sec:set-packing}

We now consider a generalization of the stochastic matching problem to hypergraphs, where each edge has size at most
$k$. Formally, the input to this \emph{stochastic $k$-set packing} problem consists of
\begin{itemize}
\item $n$ items/columns, where each item has a random profit $v_i\in
  \mathbb{R}_+$, and a random $d$-dimensional size $S_i \in \{0,1\}^d$;
  these random values and sizes are drawn from a probability
  distribution specified as part of the input. The probability
  distributions for different items are independent, as are the
  probability distributions for the value and the size for any of the
  items. Additionally, for each item, there is a set $C_i$ of at most
  $k$ coordinates such that each size vector takes positive values only
  in these coordinates; i.e., $S_i \subseteq C_i$ with probability $1$
  for each item~$i$.
\item A capacity vector $b\in \mathbb{Z}_+^d$ into which the items must
  be packed.
\end{itemize}

The parameter $k$ is called the {\em column sparsity} of the problem. The instantiation of any column (i.e., its size
and profit) is known only when it is probed. The goal is to compute an adaptive strategy of choosing items until there
is no more available capacity such that the expectation of the obtained profit is maximized.

Note that the stochastic matching problem can be modeled as a stochastic $4$-set packing problem in the following way:
we set $d = 2n$, and associate the $i^{th}$ and $(n+i)^{th}$ coordinate with the vertex $i$---the first $n$ coordinates
capture whether the vertex is free or not, and the second $n$ coordinates capture how many probes have been made
involving that vertex. Now each edge $(i,j)$ is an item whose value is $w_{ij}$; if $e_t \in \{0,1\}^d$ denotes the
indicator vector with a single $1$ in the $t^{th}$ position, then the size of the edge $(i,j)$ is either $e_i + e_j +
e_{n+i} + e_{n+j}$ (with probability $p_i$) or $e_{n+i} + e_{n+j}$ (with probability $1-p_i$). If we set the capacity
vector to be $b = (1, 1, \cdots, 1, t_1, t_2, \cdots, t_n)$, this precisely captures the stochastic matching problem.
Thus, each size vector has $\le k=4$ ones.

This stochastic $k$-set packing problem was studied (among many others) as the ``stochastic $b$-matching'' problem in
Dean~\etal~\cite{dgv05}; however the authors of that work did not consider the `column sparsity' parameter $k$ and
instead gave an $O(\sqrt{d})$-approximation algorithm for the general. Here we consider the performance of algorithms
for this problem specifically as a function of the column sparsity $k$, and prove Theorem~\ref{th:main5}.

A quick aside about ``safe'' and ``unsafe'' adaptive policies: a policy is called {\em safe} if it can include an
item only if there is {\em
  zero} probability of violating any capacity constraint. In contrast,
an {\em unsafe} policy may attempt to include an item even if there is non-zero probability of violating
capacity---however, if the random size of the item causes the capacity to be violated, then no profit is received for
the overflowing item, and moreover, no further items may be included by the policy. The model in Dean et
al.~\cite{dgv05} allowed unsafe policies, whereas we are interested (as in the previous sections) in safe policies.
However, due to the discreteness of sizes in stochastic $k$-set packing, it can be shown that our approximation
guarantee is relative to the optimal unsafe policy.

For each item $i\in [n]$ and constraint $j\in[d]$, let $\mu_i(j) :=\E[S_i(j)]$, the expected value of the $j^{th}$
coordinate in size-vector $S_i$. For each column $i\in[n]$, the coordinates $\{j\in[d] \mid \mu_i(j)>0\}$ are called
the {\em support} of column $i$. By column sparsity, the support of each column has size at most $k$. Also, let $w_i :=
\E[v_i]$, the mean profit, for each $i\in[n]$.  We now consider the natural LP relaxation for this problem, as
in~\cite{dgv05}.
\begin{linearprogram}
  {
    \label{LP:mult2}
    \maximize{\displaystyle \sum_{i=1}^n w_{i}\cdot y_{i} }
  }
   \sum_{i=1}^n \mu_i(j)\cdot y_{i} &\le b_j   &&\forall j\in [d] \label{eq:gen-lp1}\\
     y_{i} &\in [0,1]   & &\forall i \in [n] \label{eq:gen-lp2}
\end{linearprogram}

Let $y^*$ denote an optimal solution to this linear program, which in turn gives us an upper bound on any adaptive
(safe) strategy. Our rounding algorithm is a natural extension of the one for stochastic matching in
\S\ref{subsec:wt-match}. Fix a constant $\alpha\ge 1$, to be specified later. The algorithm picks a uniformly
random permutation $\pi:[n]\rightarrow [n]$ on all columns, and probes only a subset of the columns as follows. At any
point in the algorithm, column $c$ is {\em
  safe} iff there is positive residual capacity in \emph{all} the
coordinates in the support of $c$---in other words, irrespective of the instantiation of $S_c$, it can be feasibly
packed with the previously chosen columns. The algorithm inspects columns in the order of $\pi$, and whenever it is
safe to probe the next column $c\in[n]$, it does so with probability $\frac{y_c}\alpha$. Note that the algorithm skips
all columns that are unsafe at the time they appear in $\pi$.

We now prove Theorem~\ref{th:main5} by showing that this algorithm is
a $2k$-approximation. The analysis proceeds similar to that in
\S\ref{subsec:wt-match}. For any column $c\in[n]$, let
$\{\is_{c,\ell}\}_{\ell=1}^k$ denote the indicator random variables
for the event that the $\ell^{\mathrm{th}}$ constraint in the support
of $c$ is tight at the time when $c$ is considered under the random
permutation $\pi$. Note that the event ``column $c$ is safe when
considered'' is precisely $\bigwedge_{\ell=1}^k
\overline{\is_{c,\ell}}$. By a trivial union bound, the $\Pr[\, c \text{
  is safe}\,] \geq 1-\sum_{\ell=1}^k \Pr[\is_{c,\ell}]$.

\begin{lem}\label{lem:fail-prob}
  For any column $c\in[n]$ and index $\ell\in[k]$, $\Pr[\is_{c,\ell}]\le
  \frac{1}{2\alpha}$.
\end{lem}
\begin{pf}
  Let $j\in[d]$ be the $\ell^{\mathrm{th}}$ constraint in the support of
  $c$. Let $U_c^j$ denote the usage of constraint $j$, when column $c$
  is considered (according to $\pi$). Then, using argument similar to
  those used to prove Lemma~\ref{lem:match-prob}, we have
  \begin{align*}
    \E[U_c^j]&= \sum_{a=1}^n \Pr [\, \text{column $a$ appears before $c$ AND
      $a$ is probed}\, ] \cdot \mu_{a}(j), \\
    &\le  \sum_{a=1}^n \Pr [ \, \text{column $a$ appears before $c$}\,
    ] \cdot \frac{y_a}\alpha \cdot \mu_{a}(j), \\
    &= \sum_{a=1}^n \frac{y_a}{2\alpha} \cdot \mu_{a}(j), \\
    & \le \frac{b_i}{2\alpha}.
  \end{align*}
  Since $\is_{c,\ell} = \{ U_c^i\ge b_i \}$, Markov's inequality implies
  that $Pr[\is_{c,\ell}] \le \E[U_c^i]/b_i\le \frac1{2\alpha}$.
\end{pf}

Again using the trivial union bound, the probability that a particular column $c$ is safe when considered under $\pi$
is at least $1-\frac{k}{2\alpha}$, and thus the probability of actually probing $c$ is at least $\frac{y_c}{\alpha}
(1-\frac{k}{2\alpha})$. Finally, by linearity of expectations, the expected profit is at least $\frac1\alpha
(1-\frac{k}{2\alpha})\cdot\sum_{c=1}^n w_c\cdot y_c$. Setting $\alpha=k$ implies an expected profit of at least
$\frac{1}{2k}\cdot \sum_c w_cy_c$, which proves Theorem~\ref{th:main5}.

\section{Final Remarks}

An extended abstract of this paper appeared in the Proceedings of the 18th Annual European Symposium on
Algorithms~\cite{bansal10when}. The bounds presented here in \S\ref{sec:stoc-match} are slightly better than those
claimed in the extended abstract. Quite recently, Adamczyk has proved that the greedy algorithm is a $2$-approximation
for unweighted stochastic matching \cite{Adamczyk10greedy}, improving our bounds from Theorem~\ref{th:main1.5}. It
remains an open question whether the stochastic matching problem is NP-complete.

\paragraph{Acknowledgements.} We would like to thank  Aravind
Srinivasan for helpful discussions.

{
\bibliography{date}
}

\appendix
\section{Cardinality Constrained Multiple Round Stochastic Matching}
\label{sec:mult-match}

\def\mlp{\ensuremath{\mathcal{M}_C(G)}\xspace}
\def\p{\ensuremath{\mathbb{P}}\xspace}

We now consider stochastic matching with a different objective in mind; this was also defined in~\cite{c-etal}. In this
problem, we arrange for many pairs to date each other simultaneously (constrained by the fact that each person is
involved in at most one date at any time), and have $k$ days in which all these dates must happen---again, we want to
maximize the expected weight of the matched pairs.

More formally, we can probe several edges concurrently---a ``round'' may involve probing any set of edges that forms a
matching of size at most $C$. Given $k$ and $C$, the goal is to find an adaptive strategy for probing edges in rounds
such that we use at most $k$ rounds, and maximize the expected weight of matched edges during these $k$ rounds. As
before, one can probe edges involving individual $i$ at most $t_i$ times, and only if $i$ is not already matched by the
algorithm.  In this section, we give a constant-factor approximation for this problem, improving over the previously
known $O(\min\{k,C\})$-approximation~\cite{c-etal} (which only works for the unweighted case).

Our approach, as in the previous sections, is based on linear programming. The following LP captures adaptive
strategies, and hence is a relaxation of the multiple round stochastic matching problem; moreover, it can be solved in
poly-time.  Below, \mlp denotes the convex hull of all matchings in $G$ having size at most $C$.
\begin{linearprogram}
  {
    \label{LP:mult}
    \maximize{\displaystyle \sum_{(i,j)\in E} w_{ij}\cdot \sum_{h=1}^k x^h_{ij} }
  }
   \sum_{h=1}^k \,\,y^h_{ij} & \le 1 & & \forall (i,j)\in E \label{eq:mult-lp1}\\
   \sum_{j\in\partial(i)} \sum_{h=1}^k y^h_{ij} &\le t_i && \forall i\in V\label{eq:mult-lp2}\\
   y^h &\in \mlp &&\forall h\in [k] \label{eq:mult-lp3}\\
   x^h_{ij} &= p_{ij}\cdot y^h_{ij} && \forall (i,j)\in E,\,\,h\in[k]\label{eq:mult-lp4}\\
   \sum_{j\in\partial(i)} \,\,\sum_{h=1}^k x^h_{ij} &\le 1&& \forall i\in V \label{eq:mult-lp5}
\end{linearprogram}

 Since there is a linear description for \mlp, for which we can separate
 in polynomial time~\cite[Corollary~18.10a]{book/combopt/Schrijver}), the above LP can
 be solved in polynomial time using, say, the Ellipsoid algorithm.  To
 see that this LP is indeed a relaxation of the original adaptive
 problem, observe that setting $y^h_{ij}$ to be ``probability that $(ij)$
 is probed in round $h$ by the optimal strategy'' defines a feasible
 solution to the LP with objective equal to the optimal value of the
 stochastic matching instance.

Our algorithm first solves the LP to optimality and obtains solution $(x,y)$. Note that for each $h\in[k]$, using the
fact that polytope \mlp is integral and that the variables $y^h\in \mlp$, we can write $y^h$ as a convex combination of
matchings of size at most $C$; i.e., we can find matchings $\{M^h_\ell\}_\ell$ and positive values
$\{\lambda^h_\ell\}_\ell$ such that each $M^h_\ell$ is a matching in $G$ of size at most $C$ and $y^h = \sum
\lambda^h_\ell \cdot \, \chi(M^h_{\ell})$, where $\chi(M^h_{\ell})$ denotes the characteristic vector corresponding to
the edges that the are present in the matching. (See, e.g.~\cite{CarrV02}, for a polynomial-time procedure.) Fixing the
parameter $\alpha$ to a suitable value to be specified later, the algorithm does the following.

\begin{shadebox}
  \begin{OneLiners}
  \item[1.] for each round $h=1,\cdots,k$ do
    \begin{OneLiners}
    \item[a.] define the $h^{\mathrm{th}}$ matching
    $$\p^h:=\left\{
      \begin{array}{ll}
        \emptyset & \mbox{with probability } 1-\frac1\alpha\\
        M^h_\ell & \mbox{with probability } \frac{\lambda^h_\ell}{\alpha}
      \end{array}\right.
    $$
  \item[b.] Probe all edges in $\p^h$ that are safe.
  \end{OneLiners}
\end{OneLiners}
\end{shadebox}
We show that this algorithm is a $20$-approximation for $\alpha=10$, which proves Theorem~\ref{th:main4}.

As before, an edge $(i,j)\in E$ is said to be {\em safe} iff (a) $(i,j)$ has not been probed earlier, (b) neither $i$
nor $j$ is matched, and (c) neither $i$ nor $j$ has timed out.

\begin{lem}\label{lem:prob-rnd}
For any edge $(i,j)\in E$, and at round $h\in [k]$, $\Pr[\,(i,j) \text{ is safe in round } h\,]\ge 1-\frac5\alpha$.
\end{lem}
\begin{pf}
  We will show that the following three statements hold at round $h$:
  \begin{OneLiners}
  \item[i.] $\Pr[\,(i,j) \text{ has  probed} \, ]\le \frac1\alpha$.
  \item[ii.] $\Pr[\, \text{vertex $i$ is already timed out}\, ]\le \frac1\alpha$.
  \item[iii.] $\Pr[\, \text{vertex $i$ is already matched}\, ]\le \frac1\alpha$.
  \end{OneLiners}
  Since $\Pr[\, (i,j) \text{ is {\em not} safe in round } h\, ]$ is at most
  \[\Pr[\, (i,j) \text{ been probed}\, ] + \Pr[\, i \text{ matched}\,] + \Pr[\,i
  \text{ timed out}\,] + \Pr[\,j \text{ matched}\,] + \Pr[\,j \text{
    timed out}\,] \]
  by the trivial union bound, proving (i)-(iii) will prove the lemma.
  To prove~(i), observe that for any edge $e\in E$ and round $g$,
  $\Pr[\, e \text{ probed in round }g\, ]\le \Pr[e\, \in \p^g\,]=\frac1\alpha\,
  y^g_e$, and hence $\Pr[\,(i,j) \text{ probed before round }h\,]\le
  \frac1\alpha\, \sum_{g<h} y^g_e\le \frac1\alpha$, where the last
  inequality uses LP constraint~\eqref{eq:mult-lp1}.

  The proof for~(iii) is identical, using the LP
  constraint~\eqref{eq:mult-lp5}. The proof for statement~(ii) is also
  similar, though one upper bounds the expected value of the number of
  times the vertex $i$ is probed (in this step one needs to use the LP
  constraints~\eqref{eq:mult-lp2}) and then uses Markov inequality.
\end{pf}

\begin{thm}
  Setting $\alpha = 10$ gives a 20-approximation for multiple round
  stochastic matching.
\end{thm}
\begin{pf}
  Using Lemma~\ref{lem:prob-rnd}, we have for any edge $(i,j)\in E$ and
  round $h\in[k]$,
  \begin{align*}
   \Pr[\, (i,j) \text{ probed in round }h\,]
   &= \Pr[\, (i,j) \text{ safe in round }h\,]\cdot \Pr[\, (i,j)\in \p^h\mid (i,j) \text{
      safe in round }h\, ]\\
    &\ge  \left(1-\frac5\alpha\right)\cdot \Pr[\,(i,j)\in \p^h\mid (i,j) \text{ safe in round }h\,]\\
    & =  \left(1-\frac5\alpha\right)\cdot \frac{y^h_{ij}}\alpha,
  \end{align*}
  where the equality follows from the fact that events $(i,j)\in\p^h$
  and $(i,j)$ is safe in round $h$ are independent.  Thus the expected
  value accrued by the algorithm is
  $$  \sum_{e\in E} w_e \cdot \, \sum_{h=1}^k \Pr[\, e \text{ probed in
    round }h\,]\cdot p_e \ge \frac1\alpha \left(1-\frac5\alpha\right)\cdot
  \sum_{e\in E} w_e \cdot \, \sum_{h=1}^k y^h_{e}\cdot p_e,$$ which is
  $\frac1\alpha \left(1-\frac5\alpha\right)$ times the optimal LP-value.
  Setting $\alpha=10$ completes the proof.
\end{pf}

\section{Unweighted Stochastic Matching: A Greedy Algorithm}
\label{sec:unwt-greedy}

In this section we consider a greedy algorithm for the {\em
  unweighted} stochastic matching problem: in this unweighted version,
all edges have unit weight, and the goal is to maximize the expected number of matched edges. The greedy algorithm was
proposed by Chen et al.~\cite{c-etal}, and they gave an analysis proving it to be a $4$-approximation; however, the
proof was fairly involved. Here, we give a significantly simpler analysis showing an approximation guarantee of $5$.
The greedy algorithm we consider is the following:
\begin{shadebox}
  \begin{OneLiners}
  \item[1.] Let $\sigma$ denote the ordering of the edges in $E$ by
    non-increasing $p_e$-values.
  \item[2.] Consider the edges $e\in E$ in the order given by $\sigma$
    \begin{OneLiners}
    \item[a.] If edge $e$ is {\em safe} then probe it, else do not probe
      $e$.
    \end{OneLiners}
  \end{OneLiners}
\end{shadebox}
Recall that an edge is safe if neither of its endpoints have been matched or timed out. Note that the expected value of
the greedy algorithm is
\[ \alg=\sum_{e\in
  E} \Pr[\, e \text{ is matched}\,]= \sum_{e\in E} \Pr[\,e \text{ is
  probed}\,]\cdot p_e. \]

\subsection{The Analysis}

While the algorithm does not have anything to do with the linear programming relaxation we presented in the previous
section, we will use that LP for our analysis. Consider the optimal LP solution $(x^*,y^*)$, and recall that
$(x^*,y^*)$ satisfy the conditions~\eqref{eq:lp2}-\eqref{eq:lp5}. (Alternatively, use the fractional solution
$y_e^{*}:=\Pr[\, e \text{ is probed in the optimal strategy}\, ]$ and $x_e^*:=\Pr[\, e \text{ is matched in the optimal
  strategy}\,]$.)  For each $e=(i,j)\in E$, define the following three
events:

\begin{align*}
  M_e &:= \mbox{ either $i$ or $j$ is matched when $e$ is considered in
    $\sigma$}, \\
  R_e &:= \mbox{ either $i$ or $j$ has timed out when $e$ is considered
    in $\sigma$}, \mbox{ and} \\
  B_e &:= M_e \vee R_e.
\end{align*}
By the algorithm, it follows that $\Pr[\, e \mbox{ is probed}\,] = 1-\Pr[B_e]$ for all $e\in E$. So,
\begin{equation}\label{eq:unwt-basic}
  \alg = \sum_{e\in E} (1-\Pr[B_e])\, p_e \ge \sum_{e\in E} (1-\Pr[B_e])\cdot y^*_e\, p_e
\end{equation}
The following two lemmas charge the value accrued by the algorithm in two different ways to the optimal LP solution.
\begin{lem}
  \label{lem:chargeit1}
  $2\alg \ge \sum_{g\in E} \Pr[M_g]\cdot y^*_g\cdot p_g$.
\end{lem}

\begin{pf}
  In the greedy algorithm, whenever edge $e=(i,j)$ gets matched, {\em
    write} value of $\frac{y^*_f\cdot p_f}2$ on each edge $f\in
  \partial(i)\bigcup \partial(j)$. Note that the total value written
  when edge $e=(i,j)$ gets matched is at most:
  $$  \sum_{f\in \partial(i)} \frac{y^*_f\, p_f}2 + \sum_{f\in
    \partial(j)} \frac{y^*_f\, p_f}2 = \frac12 \sum_{f\in \partial(i)}
  x^*_f +\frac12 \sum_{f\in \partial(j)} x^*_f\le 1,$$
where the inequality follows from (\ref{eq:lp2}). Recall that in
  any possible execution of Greedy, an edge is matched at most once.
  Thus the expected total value written (on all edges) is at most
  $\sum_{e\in E} \Pr[\, e \text{ is matched}\,]=\alg$.

  On the other hand, whenever event $M_g$ occurs in the greedy algorithm
  (at some edge $g=(a,b)\in E$), {\em read} $\frac{y^*_g\cdot p_g}2$
  value from $g$. Consider any outcome where event $M_g$ occurs: it must
  be that either $a$ or $b$ was already matched (say via edge $e$); this
  in turn means that $\frac{y^*_g\cdot p_g}2$ value was written on edge
  $g$ at the time when $e$ got matched (since $g$ is adjacent to $e$).
  Thus the value read from an edge (at any point) is at most the value
  already written on it. Thus the expected total value read from all
  edges is $\sum_{g\in E} \Pr[M_g]\cdot \frac{y^*_g\, p_g}2\le
  \E[\mbox{total value written}]\le \alg$.
\end{pf}

\begin{lem}
  \label{lem:chargeit2}
  $2\alg \ge \sum_{g\in E} \Pr[R_g]\cdot y^*_g\cdot p_g$.
\end{lem}

\begin{pf}
  Consider the execution of the greedy algorithm, with a {\em value}
  $\alpha_e$ defined on each edge $e\in E$ (initialized to zero).
  Whenever an edge $e=(i,j)$ gets probed, do (where
  $\sigma_e$ denotes the edges in $E$ that appear after $e$ in
  $\sigma$):
  \begin{enumerate}
  \item For each $f\in \partial(i)\cap \sigma_e$, increase $\alpha_f$ by
    $\frac{y^*_f\,p_f}{2t_i}$.
  \item For each $f\in \partial(j)\cap \sigma_e$, increase $\alpha_f$ by
    $\frac{y^*_f\,p_f}{2t_j}$.
  \end{enumerate}
Let $A:=\sum_{e\in E} \alpha_e$. Note that the increase in
  $A$ when edge $e=(i,j)$ gets probed is:
  $$ \sum_{f\in \partial(i)\cap \sigma_e} \frac{y^*_f\, p_f}{2t_i} +
  \sum_{f\in \partial(j)\cap \sigma_e} \frac{y^*_f\, p_f}{2t_j} \le
  \frac{p_e}2\left( \frac1{t_i} \sum_{f\in \partial(i)\cap \sigma_e}
    y^*_f +\frac1{t_j} \sum_{f\in \partial(j)\cap \sigma_e}
    y^*_f\right)\le p_e,$$ where for the first inequality we use the
  greedy property that $p_e\ge p_f$ for all $f\in \sigma_e$ and the
  second inequality follows from (\ref{eq:lp3}). Thus the expected value
  of $A$ at the end of the greedy algorithm is $\E[A\mbox{ at the end of
    Greedy}] \le \sum_{e\in E} \Pr[\, e \text{ is probed}\,]\cdot p_e =\alg$.
  (Recall that in any possible execution of Greedy, an edge is probed at
  most once.)

  On the other hand, whenever event $R_g$ occurs in the greedy algorithm
  (at some edge $g=(a,b)\in E$), {\em read} the value $\alpha_g$ from
  $g$.  Consider any outcome where event $R_g$ occurs: it must be that
  either $a$ or $b$ was already timed out (say vertex $a$). This means
  that $t_a$ edges from $\partial(a)$ have already been probed. By the
  updates to $\alpha$-values defined above, since $g$ is adjacent to
  each edge in $\partial(a)$, the current value $\alpha_g\ge t_a\cdot
  \frac{y^*_g\,p_g}{2\, t_a}= y^*_g\,p_g/2$. So whenever $R_g$ occurs,
  the value read $\alpha_g\ge y^*_g\,p_g/2$. I.e.  the expected total
  value read is at least $\sum_{g\in E} \Pr[R_g]\cdot
  \frac{y^*_g\,p_g}{2}$.  However, the total value read is at most the
  value $A$ at the end of the greedy algorithm. This implies that $
  \sum_{g\in E} \Pr[R_g]\cdot \frac{y^*_g\, p_g}2\le \E[\mbox{total
    value read}]\le \E[A\mbox{ at the end of Greedy}] \le \alg.$
\end{pf}

\begin{proofof}{Theorem~\ref{th:main2}}
  Adding the expressions from Lemmas~\ref{lem:chargeit1}
  and~\ref{lem:chargeit2}, we get
  $$ 4\,\alg\ge \sum_{e\in E} (\Pr[M_e]+\Pr[R_e]) \cdot y^*_e\,p_e\ge
  \sum_{e\in E} \Pr[B_e] \cdot y^*_e\,p_e,$$
where the second inequality uses
  the definition $B_e=M_e\vee R_e$. Adding this
  to~\eqref{eq:unwt-basic}, we obtain $5\,\alg\ge \sum_{e\in
    E}y^*_e\cdot p_e$, which is the optimal LP objective. Thus, the
  greedy algorithm is a 5-approximation.
\end{proofof}

\end{document}